\documentclass[12pt,onecolumn]{article}

\usepackage{arxiv}
\usepackage{amsmath}
\usepackage{graphics}
\usepackage[utf8]{inputenc} 
\usepackage[T1]{fontenc}    
\usepackage{url}            
\usepackage{booktabs}       
\usepackage{amsfonts}       
\usepackage{nicefrac}       
\usepackage{microtype}      
\usepackage{lipsum}

\usepackage{amsfonts}
\usepackage{amssymb}
\usepackage{bbold}
\usepackage{booktabs}
\usepackage{ulem}
\usepackage{xr}
\usepackage[justification=centering ]{subcaption}
\usepackage{graphicx}
\usepackage{amsmath}
\usepackage[usenames,dvipsnames]{xcolor}
\usepackage{multirow}
\usepackage{lscape}
\usepackage[sorting = none, backend = bibtex, style=authoryear]{biblatex}
\usepackage{amsfonts}
\usepackage{amssymb}
\usepackage{bbold}
\usepackage{booktabs}
\usepackage{ulem}
\usepackage{xr}
\usepackage{tikz}
\usetikzlibrary{shapes}
\usepackage{soul}

\usepackage{xcolor}
\usepackage{bm}
\usepackage{amsfonts}
\usepackage{amssymb}
\usepackage{bbold}
\usepackage{hyperref}
\usepackage{capt-of}
\usepackage{float}
\usepackage{lipsum}
\usepackage{soul} 

\newcommand{\Ca}{\text{Ca}}
\newcommand{\Rey}{\text{Re}}
\newcommand{\Uw}{\vec{U}_{\mbox{\scriptsize w}}}
\newcommand{\Bq}{\text{Bq}}

\newcommand{\mum}{\mu_{\mbox{\scriptsize m}}}

\newcommand{\effi}{\mbox{n}_i}

\newcommand{\kV}{k_{\scriptsize V}}
\newcommand{\Kal}{k_{\scriptsize \alpha}}

\newcommand{\kS}{k_{\mbox{\scriptsize S}}}

\newcommand{\WS}{W_{\mbox{\scriptsize S}}}
\newcommand{\WV}{W_{\mbox{\scriptsize V}}}

\newcommand{\mus}{\mu_{\mbox{\scriptsize s}}}
\newcommand{\mud}{\mu_{\mbox{\scriptsize d}}}

\newcommand{\tl}{t_{\mbox{\tiny L}}}

\newcommand{\tlcos}{\omega}

\newcommand{\dA}{r_1}
\newcommand{\dT}{r_3}

\newcommand{\gammadot}{\dot{\gamma}}

\renewcommand{\vec}[1]{\boldsymbol{\bm{#1}}} 

\newcommand{\mur}{\mu_{\mbox{\scriptsize r}}}

\newcommand{\Caeff}{\Ca_{\mbox{\scriptsize eff}}}

\usepackage{color}
\usepackage{tikz}
\usetikzlibrary{shapes}

\definecolor{minered}{HTML}{FF0000}
\definecolor{mineblue}{HTML}{4169E1}
\definecolor{minegreen}{HTML}{228B22}
\definecolor{colorBq0}{HTML}{fff4f1}
\definecolor{colorBq10}{HTML}{fdcab6}
\definecolor{colorBq25}{HTML}{fa694a}
\definecolor{colorBq50}{HTML}{67000e}

\newcommand{\circlea}{\raisebox{0.5pt}{\tikz{\node[draw,scale=0.5,circle,fill=colorBq0](){};}}}
\newcommand{\circleb}{\raisebox{0.5pt}{\tikz{\node[draw,scale=0.5,circle,fill=colorBq10](){};}}}
\newcommand{\circlec}{\raisebox{0.5pt}{\tikz{\node[draw,scale=0.5,circle,fill=colorBq25](){};}}}
\newcommand{\circled}{\raisebox{0.5pt}{\tikz{\node[draw,scale=0.5,circle,fill=colorBq50](){};}}}

\newcommand{\trianglea}{\raisebox{0.5pt}{\tikz{\node[draw,scale=0.3,regular polygon, regular polygon sides=3,fill=colorBq0,rotate=0](){};}}}
\newcommand{\triangleb}{\raisebox{0.5pt}{\tikz{\node[draw,scale=0.3,regular polygon, regular polygon sides=3,fill=colorBq10,rotate=0](){};}}}
\newcommand{\trianglec}{\raisebox{0.5pt}{\tikz{\node[draw,scale=0.3,regular polygon, regular polygon sides=3,fill=colorBq25,rotate=0](){};}}}
\newcommand{\triangled}{\raisebox{0.5pt}{\tikz{\node[draw,scale=0.3,regular polygon, regular polygon sides=3,fill=colorBq50,rotate=0](){};}}}

\newcommand{\squarea}{\raisebox{0.5pt}{\tikz{\node[draw,scale=0.4,regular polygon, regular polygon sides=4,fill=colorBq0](){};}}}
\newcommand{\squareb}{\raisebox{0.5pt}{\tikz{\node[draw,scale=0.4,regular polygon, regular polygon sides=4,fill=colorBq10](){};}}}
\newcommand{\squarec}{\raisebox{0.5pt}{\tikz{\node[draw,scale=0.4,regular polygon, regular polygon sides=4,fill=colorBq25](){};}}}
\newcommand{\squared}{\raisebox{0.5pt}{\tikz{\node[draw,scale=0.4,regular polygon, regular polygon sides=4,fill=colorBq50](){};}}}

\newcommand{\rotsquarea}{\raisebox{0.5pt}{\tikz{\node[draw,rotate=45,scale=0.4,regular polygon, regular polygon sides=4,fill=colorBq0](){};}}}
\newcommand{\rotsquareb}{\raisebox{0.5pt}{\tikz{\node[draw,rotate=45,scale=0.4,regular polygon, regular polygon sides=4,fill=colorBq10](){};}}}
\newcommand{\rotsquarec}{\raisebox{0.5pt}{\tikz{\node[draw,rotate=45,scale=0.4,regular polygon, regular polygon sides=4,fill=colorBq25](){};}}}
\newcommand{\rotsquared}{\raisebox{0.5pt}{\tikz{\node[draw,rotate=45,scale=0.4,regular polygon, regular polygon sides=4,fill=colorBq50](){};}}}

\newcommand{\pentaa}{\raisebox{0.5pt}{\tikz{\node[draw,scale=0.4,regular polygon, regular polygon sides=5,fill=colorBq0](){};}}}
\newcommand{\pentab}{\raisebox{0.5pt}{\tikz{\node[draw,scale=0.4,regular polygon, regular polygon sides=5,fill=colorBq10](){};}}}
\newcommand{\pentac}{\raisebox{0.5pt}{\tikz{\node[draw,scale=0.4,regular polygon, regular polygon sides=5,fill=colorBq25](){};}}}
\newcommand{\pentad}{\raisebox{0.5pt}{\tikz{\node[draw,scale=0.4,regular polygon, regular polygon sides=5,fill=colorBq50](){};}}}

\newcommand{\exaa}{\raisebox{0.5pt}{\tikz{\node[draw,rotate=90,scale=0.4,regular polygon, regular polygon sides=6,fill=colorBq0](){};}}}
\newcommand{\exab}{\raisebox{0.5pt}{\tikz{\node[draw,rotate=90,scale=0.4,regular polygon, regular polygon sides=6,fill=colorBq10](){};}}}
\newcommand{\exac}{\raisebox{0.5pt}{\tikz{\node[draw,rotate=90,scale=0.4,regular polygon, regular polygon sides=6,fill=colorBq25](){};}}}
\newcommand{\exad}{\raisebox{0.5pt}{\tikz{\node[draw,rotate=90,scale=0.4,regular polygon, regular polygon sides=6,fill=colorBq50](){};}}}
\newcommand\corr[1]{\textcolor{black}{#1}}  

\newcommand{\symbolsA}{\protect\circlea, \protect\trianglea, \protect\squarea, \protect\rotsquarea, \protect\pentaa, \protect\exaa}
\newcommand{\symbolsB}{\protect\circleb, \protect\triangleb, \protect\squareb, \protect\rotsquareb, \protect\pentab, \protect\exab}
\newcommand{\symbolsC}{\protect\circlec, \protect\trianglec, \protect\squarec, \protect\rotsquarec, \protect\pentac, \protect\exac}
\newcommand{\symbolsD}{\protect\circled, \protect\triangled, \protect\squared, \protect\rotsquared, \protect\pentad, \protect\exad}

\title{\huge Suspensions of viscoelastic capsules: effect of membrane viscosity on transient dynamics}
\author{Fabio Guglietta$^{1,\dagger,*}$\And Francesca Pelusi$^{1,\ddagger}$\And Marcello Sega$^{2}$\And Othmane Aouane$^{1}$\And Jens Harting$^{1,3}$ %
\and
$^{1}$Helmholtz Institute Erlangen-Nürnberg for Renewable Energy (IEK-11), Forschungszentrum Jülich,\\ Cauerstraße 1, 91058 Erlangen, Germany
\and
$^{2}$Department of Chemical Engineering, University College London, London WC1E 7JE, United Kingdom
\and
$^{3}$Department of Chemical and Biological Engineering and Department of Physics,\\
Friedrich-Alexander-Universität Erlangen-Nürnberg, \\
Cauerstraße 1, 91058 Erlangen, Germany
\and
$^{\dagger}$Current affiliation: ~Department of Physics \& INFN, \\ Tor Vergata University of Rome, Via della Ricerca Scientifica 1, 00133, Rome, Italy
\and
$^{\ddagger}$Current affiliation: ~Istituto per le Applicazioni del Calcolo, CNR - Via dei Taurini 19, 00185 Rome, Italy\\
\and
$^{*}$\texttt{f.guglietta@fz-juelich.de}
}

\begin{document}

\maketitle

\begin{abstract}
Membrane viscosity is known to play a central role in the transient dynamics of isolated viscoelastic capsules by decreasing their deformation, inducing shape oscillations and reducing the loading time, that is, the time required to reach the steady-state deformation. However, for dense suspensions of capsules, our understanding of the influence of the membrane viscosity is minimal. In this work, we perform a systematic numerical investigation based on coupled immersed boundary -- lattice Boltzmann (IB-LB) simulations of viscoelastic spherical capsule suspensions in the non-inertial regime. We show the effect of the membrane viscosity on the transient dynamics as a function of volume fraction and capillary number. Our results indicate that the influence of membrane viscosity on both deformation and loading time strongly depends on the volume fraction in a non-trivial manner: dense suspensions with large surface viscosity are more resistant to deformation but attain loading times that are characteristic of capsules with no surface viscosity, thus opening the possibility to obtain richer combinations of mechanical features.
\end{abstract}


\section{\label{sec:introduction}Introduction}
 A capsule is formed by a liquid drop core enclosed by a thin membrane, which can be engineered with tailored mechanical properties such as strain-softening, strain-hardening and viscoelastic properties (\cite{barthes-bieselMotionDeformationElastic2016}). Capsules have emerged as a promising material for encapsulation, transportation, and sustained release of substances in various applications such as cosmetics, personal care products, self-healing paints, fire-retardant coatings, and pharmaceutical drugs (\cite{SoluteReleaseElastic2019,bahFabricationApplicationComplex2020,kim2009elastic,sun2021dual}). They are also used as a simplified model to study complex biological cells such as red blood cells numerically (\cite{zhangImmersedBoundaryLattice2007,thesis:kruger,SFFVHM18,gekle2016strongly,bacher2018antimargination}).
 The viscous component of the membrane is often disregarded when simulating the flow behaviour of red blood cells. However, microfluidic experiments have shown that, in such systems, the membrane surface viscosity is an important feature, and the interplay between the viscous and elastic contributions of the membrane is not trivial (\cite{tomaiuolo2011microfluidics,tomaiuoloStartupShapeDynamics2011a,tomaiuoloMicroconfinedFlowBehavior2016,braunmullerHydrodynamicDeformationReveals2012,pradoViscoelasticTransientConfined2015,tran-son-tayDeterminationRedBlood1984}). The mechanical and rheological properties of suspensions of purely elastic capsules have been thoroughly studied analytically (\cite{barthes1981time,barthes-bieselMotionSphericalMicrocapsule1980,barthes-bieselRoleInterfacialProperties1991,barthes-bieselTheoreticalModellingMotion1993,barthes-bieselEffectConstitutiveLaws2002}), experimentally (\cite{changExperimentalStudiesDeformation1993a,walterShearInducedDeformation2001}) and numerically (\cite{pozrikidisFiniteDeformationLiquid1995,ramanujanDeformationLiquidCapsules1998,aouaneStructureRheologySuspensions2021,pranayDepletionLayerFormation2012,karyappaDeformationElasticCapsule2014,clausenCapsuleDynamicsRheology2010,clausenRheologyMicrostructureConcentrated2011,roraiMotionElasticCapsule2015,dodsonDynamicsStrainhardeningStrainsoftening2009,KKH14,kruger2011efficient,espositoNumericalSimulationsCell2022,diazTransientResponseCapsule2000,tranModelingDeformableCapsules2020,cordascoOrbitalDriftCapsules2013a,WASH20,BAHK21,alizadbanaeiNumericalSimulationsElastic2017,kesslerSwingingTumblingElastic2008a,bagchiDynamicRheologyDilute2011}).
However, only a few studies were dedicated to understanding the effect of the capsules' membrane viscosity (\cite{barthesbiesel1985,art:yazdanibagchi13,art:lizhang19,guglietta2020effects,guglietta2021loading,guglietta2020lattice,aliLateral2022,li2021similar,diaz2001effect,zhangDynamicModeViscoelastic2020,rezghiTankTreadingDynamicsRed2022}).

In their theoretical contribution, \cite{barthesbiesel1985} performed perturbative calculations in the small-deformation limit showing that the membrane viscosity reduces the overall deformation. Concerning the loading time, that is, the time required to reach the steady-state deformation, \cite{diaz2001effect} were among the first investigating the effect of membrane viscosity on the transient dynamics using numerical simulations: using a boundary integral method they showed that, in an elongational flow, the presence of the membrane viscosity induces an increase in the loading time that is proportional to the membrane viscosity. \cite{art:yazdanibagchi13} studied the effect of the membrane viscosity on the deformation and the tank-treading frequency of a single viscoelastic capsule numerically, also observing wrinkles appearing on the surface due to the membrane viscosity.
Recently, \cite{art:lizhang19,liFinitedifferenceIntegralSchemes2020} coupled a finite difference method with the IB-LB method to simulate the effect of the viscosity at the interface. This implementation has been then employed to investigate mainly the dynamics of RBCs, highlighting the key role played by the membrane viscosity on the deformation and the associated characteristic times (\cite{guglietta2020effects,guglietta2021loading,li2021similar}) as well as on the tumbling and tank-treading dynamics (\cite{guglietta2020lattice,rezghiTankTreadingDynamicsRed2022}).

The works mentioned above investigate the effect of membrane viscosity on single capsules. However, the understanding of its effect on the suspension of capsules is still missing. To the best of our knowledge, a parametric study on the effect of membrane viscosity on such systems does not exist yet. Our contribution aims at filling this gap by focusing on generic spherical viscoelastic capsules. We present the results of a numerical investigation of the effect of membrane viscosity on suspensions of (initially spherical) viscoelastic capsules by using our coupled IB-LB implementation. 

To study the impact of membrane viscosity, quantified via the Boussinesq number $\Bq$ (see Eq.~\eqref{eq:bq}), on the deformation $D$ and loading time $\tl$, we conducted simulations using different values of $\Bq$, capillary number $\Ca$, and volume fraction $\phi$. We aim to investigate how different values of the membrane viscosity and volume fraction affect the deformation and loading time of viscoelastic capsules.

The remainder of this paper is organised as follows: in Sec.~\ref{sec:model} we present a few details on the IB-LB method (Sec.~\ref{sec:iblb_model}) and the viscoelastic membrane model (Sec.~\ref{sec:membrane_model}). \corr{In Sec.~\ref{sec:setup-parameters}, we provide details on the numerical setup and introduce the main dimensionless numbers.} Sec.~\ref{sec:results} is dedicated to the numerical results: we first show and discuss the deformation and the loading time for a single capsule (Sec.~\ref{sec:res_def_single}) and then for suspensions with different volume fraction (Sec.~\ref{sec:res_def_susp}). We finally summarise the main findings and provide some conclusions and future perspectives in Sec.~\ref{sec:summary}. 
\begin{figure}
\centering
\includegraphics[width=.9\linewidth]{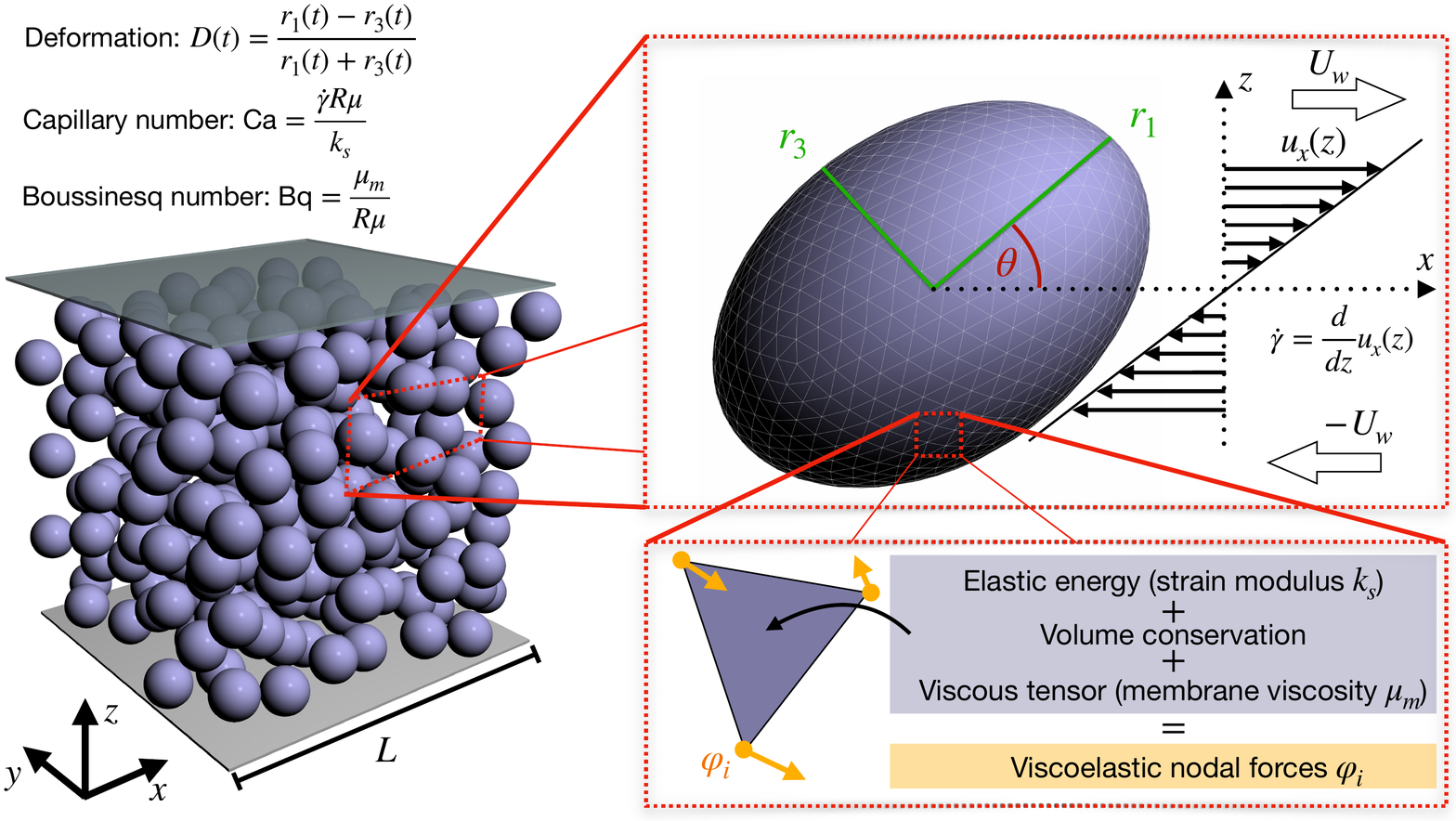}%
\caption{Sketch of the simulations performed in this work. 
Left side: 3D cubic domain with $L^3$ lattice nodes (Eulerian lattice) containing a dense suspension of viscoelastic spherical capsules with initial radius $R$. The domain is bound along the z-axis by two planar walls moving with constant speed $U_w$ in opposite directions. In this setup, we impose a simple shear flow with constant shear rate $\gammadot$. 
Top-right box: detail of a single capsule deformed under a simple shear flow. The capsules are represented using 3D triangular meshes with 2420 elements. The Taylor deformation $D$ is given by $D=(\dA-\dT)/(\dA+\dT)$, where $\dA$ and $\dT$ are the main semi-axes (green segments). The time evolution of the deformation $D(t)$ is used to evaluate the loading time $\tl$ (see Eq.~\eqref{eq:fit}). \corr{The inclination angle $\theta$ is the angle that $r_1$ forms with the flow direction ($x$-axis).}
Bottom-right box: on each triangular element, the viscoelastic forces are computed and distributed to the vertices. 
These forces are coupled to the fluid via the immersed boundary (IB) method and the fluid dynamics is simulated using the lattice Boltzmann (LB) method (see Sec.~\ref{sec:model}).
}\label{fig:sketch}%
\end{figure}

\section{Numerical model}\label{sec:model}
We simulate the dynamics of the capsules and the surrounding fluid using the coupled IB-LB method. In a nutshell, the IB method uses a triangulated mesh of Lagrangian points as support to compute forces that are then used to impose the correct space and time-dependent boundary conditions on the fluid, which is simulated using the LB method. The IB-LB method provides a two-way coupling: the boundary surface deforms due to the fluid flow, and the fluid local momentum balance is changed due to the viscoelastic forces exerted by the boundary surface. Boundary surface forces
comprise membrane elasticity, membrane viscosity, a volume-conserving regularization term, and a repulsive force to prevent capsules from penetrating each other. Details are reported below.

\subsection{The immersed boundary - lattice Boltzmann method}\label{sec:iblb_model}
The LB method solves numerically a discretised version of the Boltzmann transport equation for the particle populations $\effi$, representing the  
probability density function of fluid molecules moving with a discrete velocity $\vec{c}_i$ at position $\vec{x}$ on the lattice and at time $t$ (\cite{benzi1992lattice}). The solution to the Navier-Stokes equations emerges from the transport equation via the calculation of the moments of the particle distribution and the appropriate Chapman-Enskog analysis (\cite{chapman1990mathematical}).

The evolution of the functions $\effi$ provided by the LB equation  is
\begin{equation}\label{LBMEQ}
\effi(\vec{x}+\vec{c}_i\Delta t, t+ \Delta t) - \effi(\vec{x}, t) = \Omega_i + S_i\ ,
\end{equation}
where $\Delta t$ is the discrete time step, $\Omega_i$ represents the collision operator and $S_i$ is a source term proportional to the acting external forces $\vec{F}$ (such as membrane forces, see Sec.~\ref{sec:membrane_model}) that is implemented following \cite{PhysRevE.65.046308}: 
\begin{equation}
S_i(\vec{x},t)=\left(1-\frac{\Delta t}{2\tau}\right)\frac{w_i}{c_s^2}\left[\left(\frac{\vec{c}_i\cdot \vec{u}}{c_s^2}+1\right)\vec{c}_i-\vec{u}\right]\cdot\vec{F}\ ,
\end{equation}
Here, $\tau$ is the relaxation time, i.e., the time the functions $\effi$ take to reach the equilibrium distribution $\effi^{(\mbox{\tiny eq})}$ which is given by (\cite{qian1992lattice})
\begin{equation}
\effi^{(\mbox{\tiny eq})}(\vec{x},t)= w_i\rho\left(1+\frac{\vec{u}\cdot\vec{c}_i}{c_s^2}+\frac{(\vec{u}\cdot\vec{c}_i)^2}{2c_s^4}-\frac{\vec{u}\cdot\vec{u}}{c_s^2}\right)\ ,
\end{equation}
with $c_s=\Delta x/\Delta t\sqrt{3}$ being the speed of sound, $\Delta x$ the lattice spacing and $w_i$ suitable weights. In the D3Q19 scheme used in this work, $w_0=1/3$, $w_{1-6}=1/18$, $w_{7-18}=1/36$.  
We implement the Bhatnagar-Gross-Krook collision operator (\cite{qian1992lattice}) 
\begin{equation}\label{bgk}
\Omega_i= -\frac{\Delta t}{\tau}\left(\effi(\vec{x}, t) - \effi^{(\mbox{\tiny eq})}(\vec{x}, t)\right) \ .
\end{equation}
The Chapman-Enskog analysis provides the bridge between the LB and the Navier-Stokes equations by linking the relaxation time $\tau$ to the fluid transport coefficients, for example the dynamic viscosity 
\begin{equation}
    \mu= \rho c_s^2\left(\tau-\frac{\Delta t}{2}\right)\ .
\end{equation}
The functions $\effi$ are then used to compute the hydrodynamic density ($\rho$) and velocity ($\vec{u}$) fields of the fluid as
\begin{equation}
\rho(\vec{x}, t) = \sum_{i} \effi(\vec{x}, t)\; , \qquad\qquad \rho\vec{u}(\vec{x}, t) = \sum_{i} \vec{c}_i \effi(\vec{x}, t) + \frac{\vec{F}\Delta t}{2}\ .  
\end{equation}

The coupling between the fluid and the viscoelastic membrane is accounted through the IB method. The membrane is represented by a set of Lagrangian nodes linked to build a 3D triangular mesh (see Fig.~\ref{fig:sketch}). The idea is to interpolate the fluid (Eulerian) velocity ($\vec{u}$) to compute the nodal (Lagrangian) velocity ($\vec{\dot{r}}$) and to spread the nodal force ($\vec{\varphi}$) to find the force \corr{density} acting on the fluid ($\vec{F}$). Such interpolations are given by the following equations (\cite{book:kruger,peskin2002immersed}):
\begin{equation}\label{eq:ibm_force_d}
\vec{F}(\vec{x},t) =  \sum_i\vec{\varphi}_i(t)\Delta(\vec{r}_i-\vec{x})\ ,\qquad\qquad
\dot{\vec{r}}_i(t) = \sum_{\vec{x}} \vec{u}(\vec{x},t)\Delta(\vec{r}_i-\vec{x})\Delta x^3\ ,
\end{equation}
where $\Delta$ is a discretised approximation of a Dirac delta function which can be factorised as the product of three interpolation stencils $\Delta(\vec{x}) = \phi(x)\phi(y)\phi(z)\corr{/\Delta x^3}$. In this work, we use the two-point interpolation stencil
\begin{equation}
\phi_2(x) = 
    \begin{cases}
         1 - |x| & \mbox{for } 0\le |x|\le 1\ ,\\
         0 & \mbox{elsewhere}\ . \\
    \end{cases}
\end{equation}

\subsection{Membrane model}\label{sec:membrane_model}
\subsubsection{Elastic model} We use the Skalak model to account for the membrane elasticity (\cite{art:skalaketal73}). Here, the elastic free energy is given by 
\begin{equation}\label{eq:skalak}
\WS = \sum_j A_j\left[ \frac{\kS}{12}\left(I_{1,j}^2+2I_{1,j}-2I_{2,j}\right) +  \frac{\Kal}{12} I_{2,j}^2\right]\ ,
\end{equation}
where $A_j$ is the area of the $j-$th triangular element of the mesh, $\kS$ and $\Kal$ are the elastic shear and dilatational moduli (we restrict ourselves to $\Kal = \kS$.), respectively, $I_{1,j} = \lambda_{1,j}^2+\lambda_{2,j}^2-2$ and $I_{2,j} = \lambda_{1,j}^2\lambda_{2,j}^2-1$ are the strain invariants for the $j$-th triangular element, with $ \lambda_{1,j}$ and $ \lambda_{2,j}$ being the principal stretch ratios of the triangle (\cite{art:skalaketal73,KKH14}). The free energy $\WS^{(j)}$ computed on the $j-$th element is used to compute the force on its three vertices: we can write the force acting on the $i-$th node with coordinates $\vec{x}_i$ as
\begin{equation}\label{eq:from_w_to_f}
    \vec{\varphi}_i = -\frac{\partial \WS^{(j)}}{\vec{x}_i}\ .
\end{equation}

\subsubsection{Viscous model} 
The membrane viscosity can be implemented through the incorporation of the viscous stress tensor given by 
\begin{equation}\label{eq:bq-scriven-law}
\pmb{\tau}_\nu =  \mus \left(2\vec{e} -\mbox{tr}(\vec{e})\vec{P}\right) + \mud \mbox{tr}(\vec{e})\vec{P}=2\mum\vec{e}\ ,
\end{equation}
where $\mus$ and $\mud$ are, respectively, the shear and dilatational membrane viscosity (in order to reduce the number of parameters, we consider $\mus=\mud=\mum$, and we will only refer to the membrane viscosity $\mum$ (\cite{barthesbiesel1985})), $\vec{P}$ is the projector tensor to the 2D surface, and 
\begin{equation}\label{eq:strain}
    \vec{e} =\frac{1}{2}\left\{\boldsymbol{P}\cdot\left[\left(\boldsymbol{\nabla}^{\boldsymbol{S}}\mathbf{u}^{\boldsymbol{S}}\right)+\left(\boldsymbol{\nabla}^{\boldsymbol{S}}\mathbf{u}^{\boldsymbol{S}}\right)^{\dagger}\right]\cdot\boldsymbol{P}\right\}
\end{equation}
is the surface rate of strain.
In Eq.~\eqref{eq:strain}, the superscript $\boldsymbol{S}$ identifies the surface projection of the  gradient operator ($\boldsymbol{\nabla}^{\boldsymbol{S}}$) and local membrane velocity ($\vec{u}^{\boldsymbol{S}}$) (\cite{art:lizhang19}).
By following \cite{art:lizhang19}, we employ the standard linear solid model to compute $\pmb{\tau}_\nu$.
We evaluate the stress tensor $\vec{\tau}_\nu^{(j)}$ on each triangular element $j$ (i.e., we rotate the triangular element on the $xy$-plane), and we then compute the force on its vertices $i$ as
\begin{equation}
    \vec{\varphi}_i(x,y) = A_j\vec{\mathcal{P}}^{(j)}\cdot\vec{\nabla}N_i\ ,
\end{equation}
where $N_i(x,y)=a_ix+b_iy+c_i$ are the linear shape functions, the tensor $\vec{\mathcal{P}}^{(j)}=\left[\vec{\tau}_\nu\cdot(\vec{\mathcal{F}}^{-1})^T\right]^{(j)}$, with $(\vec{\mathcal{F}}^{-1})^T$ being the transpose of the inverse of the deformation gradient tensor $\vec{\mathcal{F}}$ (\cite{thesis:kruger,art:lizhang19,guglietta2020effects}).

\subsubsection{Volume conservation} In addition to the previous two contributions to the nodal force, we also impose the volume conservation by adding another term to the elastic free energy given in Eq.~\eqref{eq:skalak}:
\begin{equation}
    \WV = \kV \frac{(V-V_0)^2}{2V_0}\ .
\end{equation}
$\kV$ is an artificial modulus tuning the strength of the volume conservation, $V$ is the total volume of the capsule (the subscript $0$ refers to the volume at rest, i.e., $V_0=4\pi R^3/3$) (\cite{thesis:kruger,aouaneStructureRheologySuspensions2021}). The nodal force is then computed in the same way as for the elastic model (Eq.~\eqref{eq:from_w_to_f}).

\subsubsection{Capsule-capsule repulsion} Finally, to avoid capsules penetrating each others, we introduce a force 
\begin{equation} 
    \vec{\varphi}_{ij} = 
    \begin{cases}
             \bar{\epsilon}\left[\left(\frac{\Delta x}{d_{ij}}\right)^2-\left(\frac{\Delta x}{\delta_0}\right)^2\right]\hat{\vec{d}}_{ij} & \mbox{if } d_{ij}<\delta_0\ ,\\
              0 & \mbox{if } d_{ij}\ge\delta_0\ , \\
    \end{cases}
\end{equation}
acting on nodes $i$ and $j$ belonging to two different capsules, where $d_{ij}$ is the distance between nodes $i$ and $j$, $\hat{\vec{d}}_{ij}=\frac{\vec{d}_{ij}}{d_{ij}}$ is the unit vector connecting them, $\delta_0$ is the interaction range and $\bar{\epsilon}\approx 100/3 \kS$. The choice of the parameter $\bar{\epsilon}$ is as such that the macroscopic behaviour of the suspension is not affected by this additional nodal force contribution (\cite{aouaneStructureRheologySuspensions2021} provide further details). 

\subsection{Membrane geometry}\label{sec:geometry}
The information on the geometry of the capsules is retrieved from the inertia tensor, which is defined  by (\cite{thesis:kruger,ramanujanDeformationLiquidCapsules1998})
\begin{equation}
    \mathcal{I}_{\alpha\beta} = \frac{\rho_p}{5}\sum_iA_i (\vec{r}_i^2\delta_{\alpha\beta}-r_{i\alpha}r_{i\beta})r_{i\gamma}n_{i\gamma}\ .
\end{equation}
Here, $\rho_p$ is the density of the particle (in our case, $\rho_p=1$), $\vec{r}_i$ is a vector pointing form the centre of mass of the capsule to the centroid of face $i$. $A_i$ and  $\vec{n}_i$ are the area and the unit normal of the face $i$, respectively. We now consider the inertia ellipsoid, i.e., the equivalent ellipsoid with the same inertia tensor $\vec{\mathcal{I}}$. The three eigenvalues ($\mathcal{I}_1$, $\mathcal{I}_2$ and $\mathcal{I}_3$) can be used to compute the lengths of the three semi-axes of the ellipsoid with density $\rho_p$ and volume $V$ (\cite{thesis:kruger,ramanujanDeformationLiquidCapsules1998}):
\begin{align}\label{eq:semiaxes}
    r_1 = \sqrt{\frac{5(\mathcal{I}_2+\mathcal{I}_3-\mathcal{I}_1)}{2\rho_p V}}\ , \\
    r_2 = \sqrt{\frac{5(\mathcal{I}_1+\mathcal{I}_3-\mathcal{I}_2)}{2\rho_p V}}\ , \\
    r_3 = \sqrt{\frac{5(\mathcal{I}_1+\mathcal{I}_2-\mathcal{I}_3)}{2\rho_p V}}\ ,
\end{align}
with $r_1\ge r_2 \ge r_3$. By comparing with Fig.~\ref{fig:sketch}, $\dA$ and $\dT$ are the longest and shortest radii in the shear plane (respectively), while $r_2$ is the radius directed along the vorticity direction (y-axis). 

Once we know the length of the two main semi-axes $r_1$ and $r_3$, we can evaluate the deformation index 
\begin{equation}\label{eq:def}
    D(t) = \frac{\dA(t)-\dT(t)}{\dA(t)+\dT(t)}\ ,
\end{equation}
which is equal to zero when the spherical capsule is not deformed (i.e., $\dA=\dT$).

\corr{Finally, the inclination angle $\theta$ (see Fig.~\ref{fig:sketch}) is the angle that the longest radius $r_1$ forms with the flow direction ($x-$axis).}

\begin{figure}
\centering
\includegraphics[width=.6\linewidth]{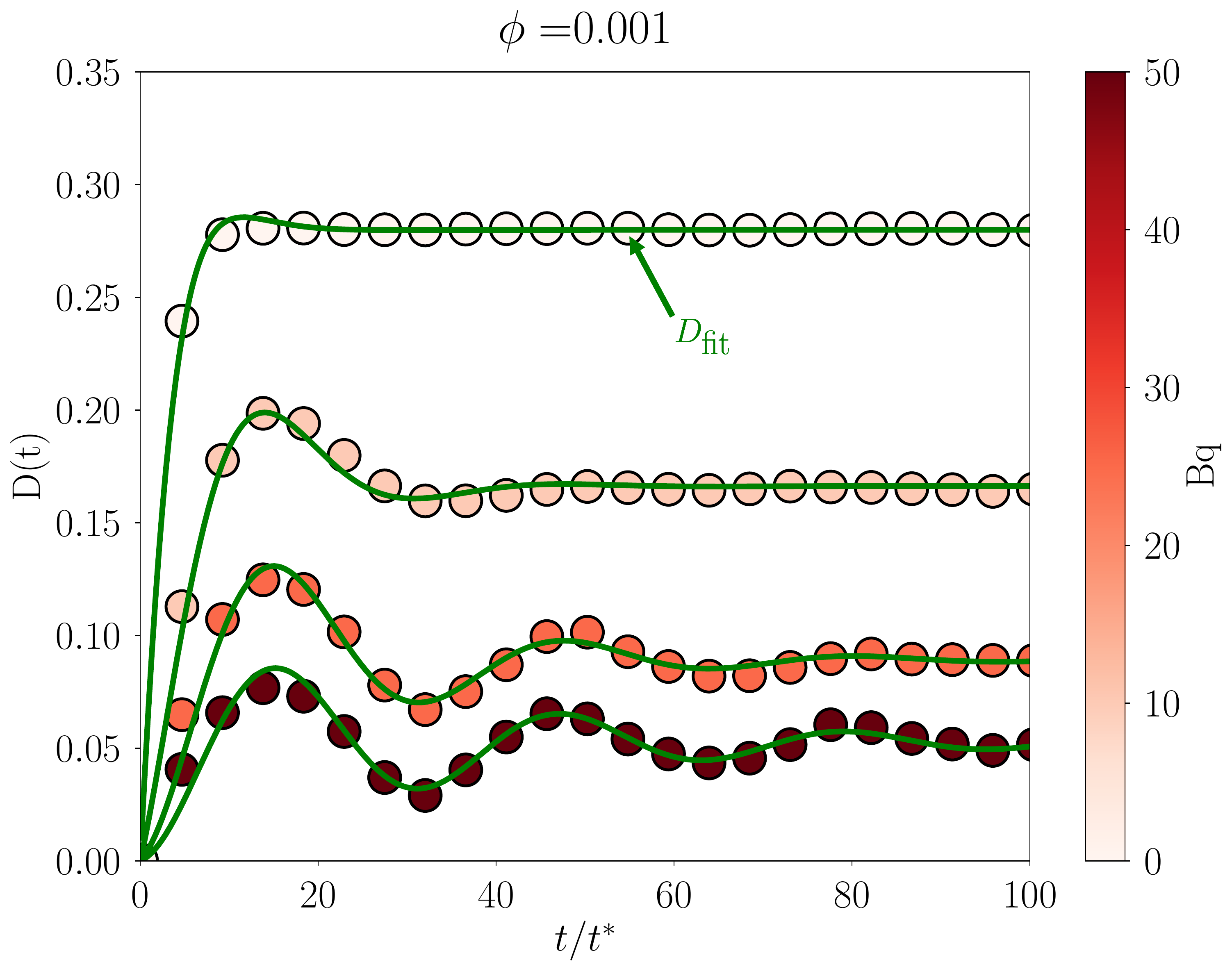}%
\caption{Deformation of the single capsule ($\phi=0.001$) as a function of  $t/t^*$ for $\Ca=0.2$ and different values of the $\Bq$ ($\Bq=0$ (\protect\circlea), $\Bq=10$ (\protect\circleb), $\Bq=25$ (\protect\circlec), $\Bq=50$ (\protect\circled)). The solid lines represent the best fit to Eq.~\eqref{eq:fit}.}\label{fig:d_vs_t-single}%
\end{figure}

\begin{figure}
\centering
\includegraphics[width=.8\linewidth]{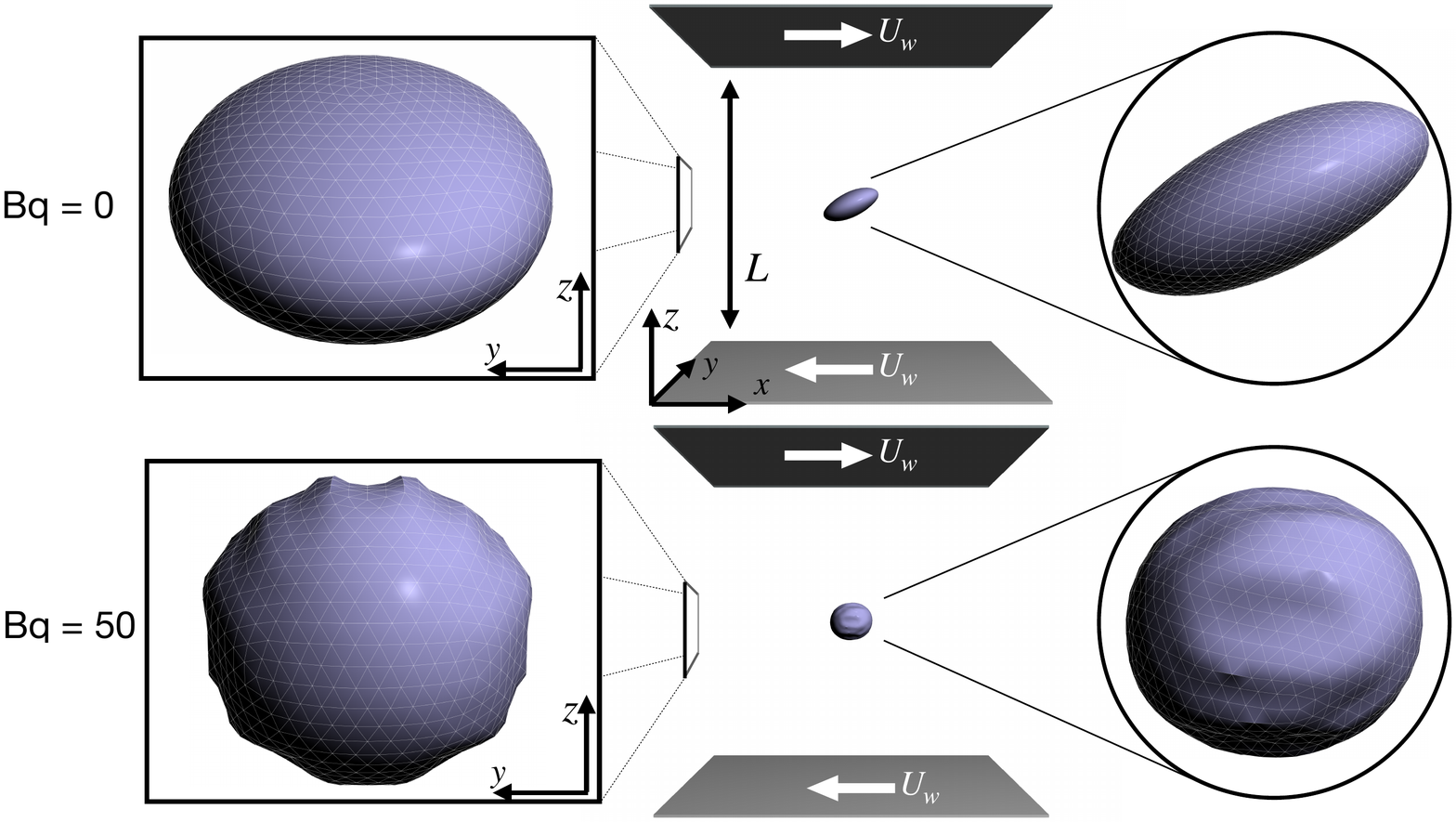}   %
\caption{Steady-state configurations for a single capsule ($\phi=0.001$) under shear flow with $\Ca=0.5$. Top panel: single capsule configuration with $\Bq=0$.  Bottom panel: single capsule configuration with $\Bq=50$. Left part: side view in the $yz$-plane. Right part: side view in the $xz$-plane.}\label{fig:sketch-single}%
\end{figure}

\section{\corr{Simulation setup and physical parameters}}\label{sec:setup-parameters}

The numerical setup consists of a cubic Eulerian domain with $L^3$ lattice nodes, where $L=128\,\Delta x$. The domain is bound along the z-axis by two planar walls at which we impose a constant velocity $U_w$ to generate a simple shear flow with constant shear rate $\gammadot$ (see Fig.~\ref{fig:sketch}).
The viscoelastic capsules have an initial radius $R=8\,\Delta x$, and the corresponding mesh is made of 2420 triangular elements. Each capsule is initialised as a rigid sphere in order to start the simulation with zero stress and deformation of the surface. Furthermore, the distance between the surfaces of the capsules cannot be less than one lattice spacing.

Several dimensionless numbers may play a role in describing the dynamics of the system. First of all, the Reynolds number 
\begin{equation}
    \Rey=\frac{\gammadot R^2\rho}{\mu}\  
\end{equation}
gives the balance between inertial and viscous forces. We chose $\Rey$ small enough ($\Rey\sim 10^{-2}$) to neglect inertial effects. The capillary number
\begin{equation}\label{eq:ca}
    \Ca = \frac{\gammadot R \mu}{k_s}\ 
\end{equation}
measures instead the importance of the viscosity of the fluid with respect to the elasticity of the membrane: we chose the range of $\Ca$ in order to work as close as possible to the small-deformation regime, avoiding strongly non-linear effects ($\Ca\in[0.05,1]$). \corr{In this paper, we have purposefully chosen to equate the elastic dilatational modulus ($\Kal$) and the elastic shear modulus ($\kS$). This decision has been made to decrease the complexity of parameters within our simulations, aligning with our primary aim of centring the study on the effects of surface viscosity.} The dimensionless number accounting for the membrane viscosity $\mum$ is the Boussinesq number
\begin{equation}\label{eq:bq}
    \Bq = \frac{\mum}{\mu R}\ ,
\end{equation}
which describes the importance of the membrane viscosity with respect to the
fluid viscosity (in this work, we consider the range $\Bq\in[0,50]$). Note that $\mum$ describes the viscosity of a 2D membrane: for this reason, it is measured in [m Pa s], while the fluid viscosity is given in [Pa s]. Finally, for dense suspensions, it is important to define the volume fraction
\begin{equation}\label{eq:phi}
    \phi = \frac{\sum_iV_i}{L^3}\ ,
\end{equation}
which ranges in $\phi\in[0.001,0.4]$ (i.e., from 1 to 400 capsules). In Eq.~\eqref{eq:phi}, $\sum_i V_i$ coincides with the total volume occupied by the viscoelastic spheres.
The computational time is normalised with the capillary time as
\begin{equation}\label{eq:tstar}
    t^* = \frac{R \mu}{k_s}\ .
\end{equation}
\corr{Note that, in this work, the viscosity ratio is unity, meaning that the viscosity of the fluid inside the capsules is equivalent to that of the fluid outside.} The main quantities mentioned above are also summarised in Tab.~\ref{tab:values}.

\corr{We also briefly mention the roles played by the membrane viscosity and the internal fluid one. Indeed, in order to simulate the effect of membrane viscosity,  \cite{keller1982motion} were the first to propose an effective viscosity ratio that is the sum of the viscosity ratio $\lambda$ and a term which accounts for the dissipation due to the membrane viscosity. However, some recent studies showed that while the qualitative effect of both kinds of viscosity is similar, they quantitatively show different behaviours~\cite{guglietta2020lattice,li2021similar,matteoli2021impact,noguchi2005dynamics,noguchi2007swinging}. We decided to keep the viscosity ratio $\lambda=1$ to focus on the effect of membrane viscosity only and avoid enlarging the already wide space of parameters.}

\begin{table}
  \begin{center}
\def~{\hphantom{0}}
  \begin{tabular}{ccc}
$L$\qquad    &(length of the domain)            & $128\,\Delta x$ \\ \hline
$R$\qquad    &(radius of the spherical capsule) & $8\,\Delta x$   \\ \hline
$\Rey$\qquad &(Reynolds number)                 & $\sim 0.01$     \\ \hline
$\Ca$\qquad  &(Capillary number)                & 0.05 - 1.0      \\ \hline
$\Bq$\qquad  &(Boussinesq number)               & 0 - 50          \\ \hline
$\phi$\qquad &(Volume fraction)                 & 0.001 - 0.4     \\ \hline
\end{tabular}
  \caption{Simulation parameters \corr{in lattice units.}}\label{tab:values}
  \end{center}
\end{table}

Intending to study and quantify the transient deformation of viscoelastic capsules, we use the solution of a damped oscillator to describe the deformation behaviour as a function of \corr{the dimensionless} time: 
\corr{\begin{equation}\label{eq:fit}
    D_{\mbox{\tiny fit}}\left( \frac{t}{t^*}\right)=\bar{D}\left[1-\exp\left(-\frac{t}{t^*\tl}\right)\cos{\left(\omega \frac{t}{t^*}\right)}\right]\ 
\end{equation}}
where $\bar{D}$ represents the steady-state value of the deformation, $\tl$ is the \corr{dimensionless} loading time (i.e., the time the capsule takes to deform) and $\tlcos$ coincides with the \corr{dimensionless} frequency of the deformation oscillations. To show how Eq.~\eqref{eq:fit} fits data from numerical simulations, in Fig.~\ref{fig:d_vs_t-single} we report the measured deformation $D$ as a function of the dimensionless time $t/t^*$ for the single capsule case. Different colours correspond to different values of $\Bq$, while all data refer to the case with $\Ca=0.2$. Fig.~\ref{fig:d_vs_t-single} shows an excellent agreement between $D_{\mbox{\tiny fit}}(t)$ (solid lines) and the numerical simulations (circles), confirming that Eq.~\eqref{eq:fit} is a suitable estimate for the dynamical observables $\tl$ and $\tlcos$.

\corr{Concerning the choice of making time dimensionless, there are mainly two choices: either using the shear rate $\dot{\gamma}$ or the capillary time $t^*$ (\cite{diazTransientResponseCapsule2000,maffettone1998equation,barthes-bieselMotionDeformationElastic2016}). In particular, \cite{barthes-bieselMotionDeformationElastic2016} considered a capsule with membrane viscosity under simple shear flow, and they observed that the response (loading) time made dimensionless via the intrinsic time decreases with the capillary number. Moreover, \cite{guglietta2020effects,guglietta2021loading} studied the transient dynamics of red blood cells under simple shear flow and in order to compare their numerical results against experiments, they reported the characteristic loading and relaxation times (in $[s]$ on the y-axis) as functions of the shear rate $\dot{\gamma}$ (in $[s^{-1}]$ on the x-axis). We therefore decided to take this as an example, and to normalise both x- and y-axis with the capillary time $t^*$, thus obtaining the dimensionless loading time $t_L$ as a function of the capillary number $\Ca$.}

\section{Results}\label{sec:results}
\corr{In this section,} we show the numerical results concerning the deformation $D$ and the loading time $\tl$ of both a single spherical capsule (Sec.~\ref{sec:res_def_single}) and a suspension of particles (Sec.~\ref{sec:res_def_susp}). 


\subsection{Single capsule}\label{sec:res_def_single}
\begin{figure}
\centering
    \includegraphics[width=1.\linewidth]{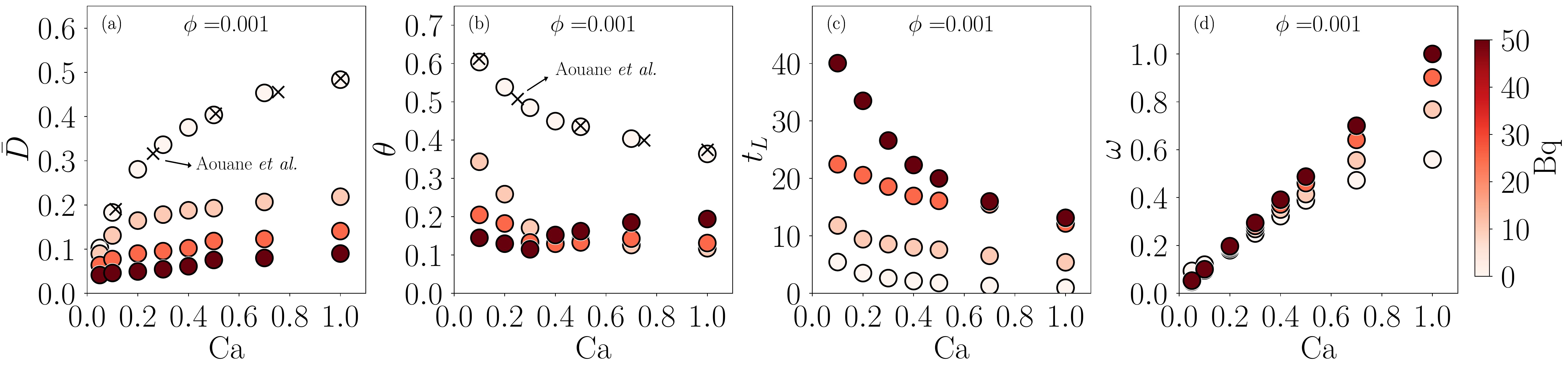} 
\caption{\corr{Data corresponding to the single capsule case ($\phi=0.001$) for different values of the $\Bq$ ($\Bq=0$ (\protect\circlea), $\Bq=10$ (\protect\circleb), $\Bq=25$ (\protect\circlec), $\Bq=50$ (\protect\circled)).
 Panel (a): steady-state deformation $\bar{D}$ as a function of the capillary number $\Ca$, where black crosses represent data from \cite{aouaneStructureRheologySuspensions2021}.
 Panel (b): inclination angle $\theta$ as a function of the capillary number $\Ca$.
 Panel (c): loading time $\tl$ as a function of the capillary number $\Ca$.
 Panel (d): frequency $\tlcos$ as a function of the capillary number $\Ca$.}
 }\label{fig:d_and_t_vs_ca-single}%
\end{figure}
\begin{figure}
\centering
\includegraphics[width=.6\linewidth]{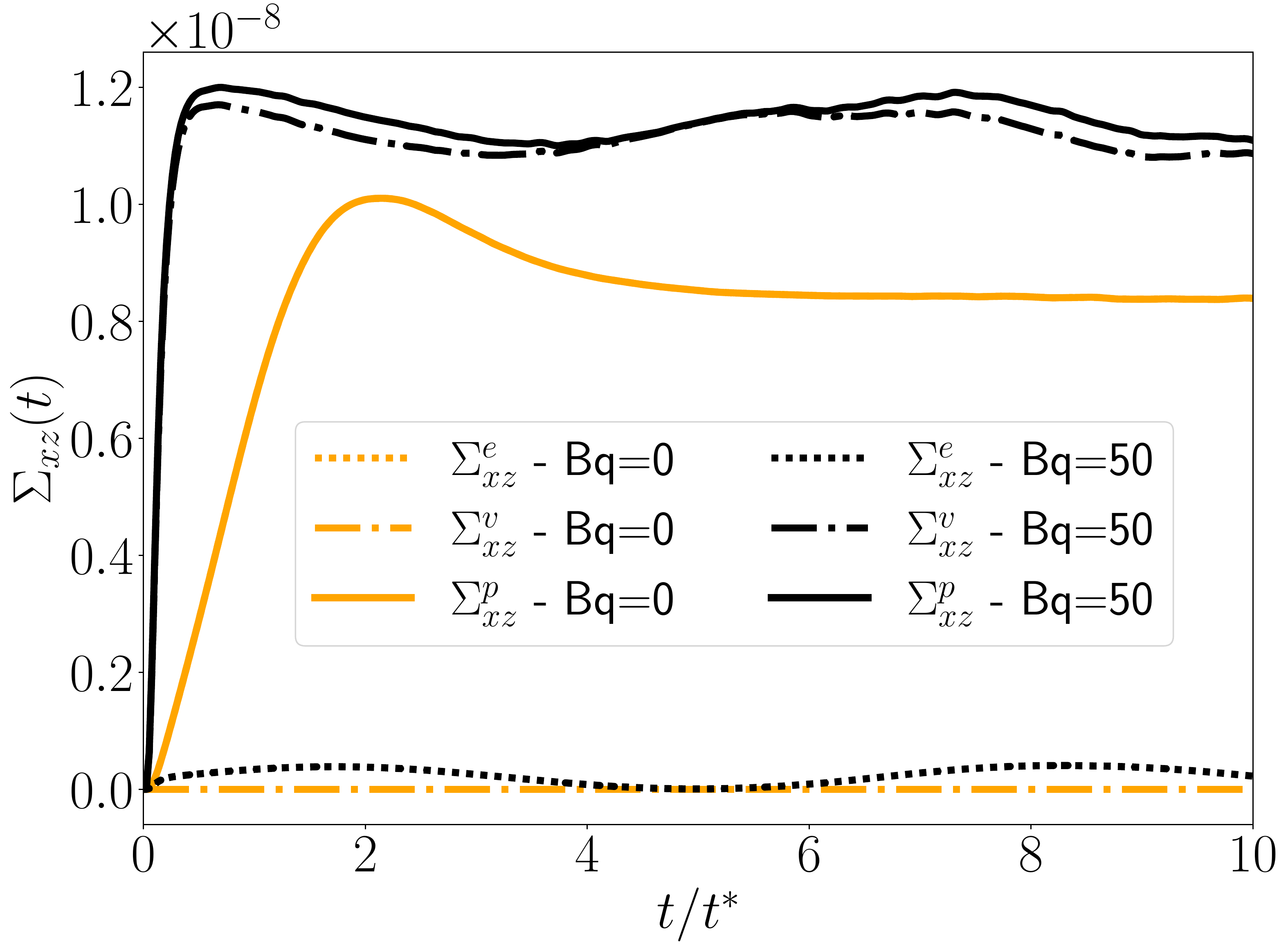}%
\caption{\corr{Time evolution of the xz-components of the particle stress $\Sigma^p_{xz}$ for a single capsule, for Bq=0 (orange lines) and Bq=50 (black lines). Dotted and dash-dotted lines represent the xz-component of the elastic ($\Sigma^e_{xz}$) and viscous ($\Sigma^v_{xz}$) contributions of the stress, respectively; solid lines represent the particle stress $\Sigma^p_{xz} =(\Sigma^e+\Sigma^v)_{xz}$.}}\label{fig:sigma_vs_t-single}%
\end{figure}
\begin{figure}
\includegraphics[width=1.\linewidth]{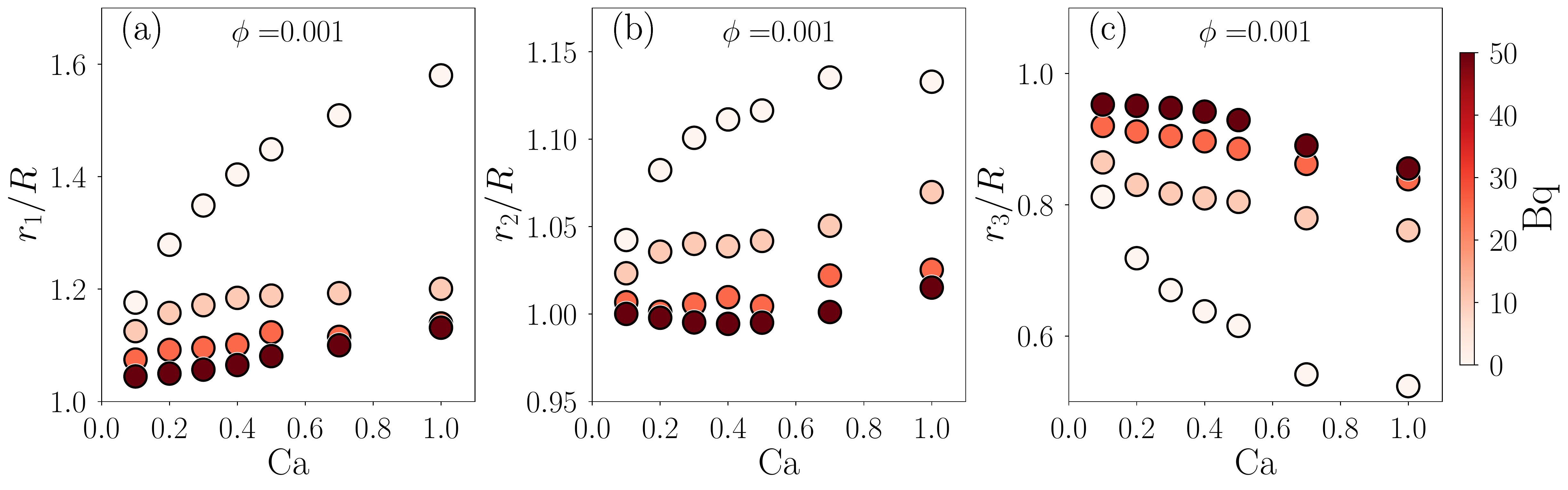}
\caption{Three main radii of the single capsule ($\phi=0.001$) as a function of $\Ca$, for different values of $\Bq$ ($\Bq=0$ (\protect\circlea), $\Bq=10$ (\protect\circleb), $\Bq=25$ (\protect\circlec), $\Bq=50$ (\protect\circled)), normalised to the capsule radius at rest, $R$.}\label{fig:r_vs_ca-single}%
\end{figure}
\begin{figure}
  \includegraphics[width=1.\linewidth]{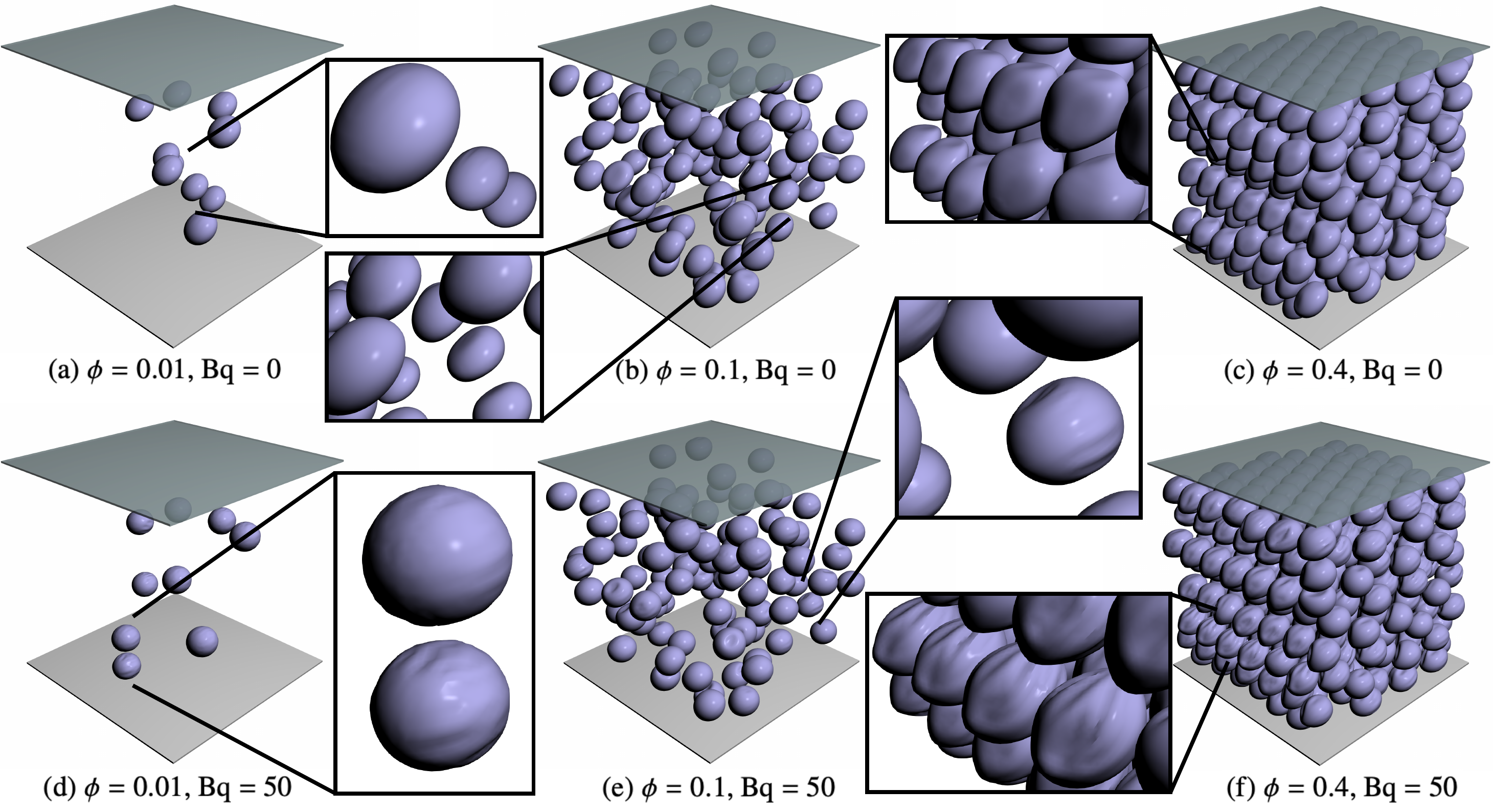}
\caption{\corr{Snapshots of the suspensions. The configurations shown correspond to $\Ca=0.1$ and $\Bq=0$ (top panels, (a)-(c)) and $\Bq=50$ (bottom panels, (d)-(f)).}}\label{fig:sketch-susp}%
\end{figure}

In this section, we report the numerical results for the deformation and loading time of a single capsule ($\phi=0.001$) under shear flow, which will serve as a reference for the next section, where suspensions of capsules are considered. 
Fig.~\ref{fig:sketch-single} shows the steady-state configuration of a capsule under shear flow with $\Ca=0.5$, for two values of the $\Bq$ ($\Bq=0$, top panels; $\Bq=50$, bottom panels). The capsule is initialised in the middle of the channel; white arrows represent the velocity of the walls $\Uw$. Left and right parts of Fig.~\ref{fig:sketch-single} show side views in the $yz$- and $xz$-plane, respectively. 
Fig.~\ref{fig:sketch-single} shows that some wrinkles appear on the surface when $\Bq$ increases. These results agree with what was observed by \cite{art:yazdanibagchi13}. 
\corr{It should be noted that the introduction of a bending energy into the membrane model can potentially inhibit the emergence of these wrinkles, as discussed in details in \cite{art:yazdanibagchi13}.}
In the case of a purely elastic capsule ($\Bq=0$), no wrinkles appear if the capillary number is large enough ($\Ca\gtrapprox 0.1$), but some of them do appear when the capillary number is small ($\Ca=0.05$). We emphasise that these wrinkles are not a numerical artefact, as they have also been observed in experiments (\cite{walterShearInducedDeformation2001,unverfehrt2015deformation}) and analytically studied (\cite{finkenWrinklingMicrocapsulesShear2006}).

In Fig.~\ref{fig:d_and_t_vs_ca-single}(a), we show the steady-state value of the deformation $\bar{D}$ as a function of the capillary number $\Ca$ for different values of $\Bq$ (the darker the colour, the higher the value of $\Bq$). We also report results from \cite{aouaneStructureRheologySuspensions2021} (black crosses), as a benchmark of our implementation, which corresponds to a case without membrane viscosity. 
 
Fig.\ref{fig:d_and_t_vs_ca-single}(a) shows that the effect of increasing $\Bq$ is to decrease the deformation, a trend that has been previously observed in other works (\cite{art:yazdanibagchi13,art:lizhang19,guglietta2020effects,guglietta2021loading}). This can be explained by an energetic argument: for a fixed value of the elastic modulus $\kS$ and a given intensity of the shear rate $\gammadot$ (i.e., for the same value of the capillary number $\Ca$), the energy injected into the system is the same. However, the simple shear flow can be split into two contributions, accounting for the rotation and the elongation of the capsule, respectively: 
\begin{equation}
\vec{\nabla}\vec{u} = 
\left(\begin{matrix}
0 & \dot{\gamma} \\
0 & 0
\end{matrix}\right)=
\left(\begin{matrix}
0 & \frac{\dot{\gamma}}{2} \\
\frac{\dot{\gamma}}{2} & 0
\end{matrix}\right) + \left(\begin{matrix}
0 & \frac{\dot{\gamma}}{2}  \\
-\frac{\dot{\gamma}}{2}  & 0
\end{matrix}\right)\; 
\end{equation}

This means that the energy injected by the applied shear flow not only contributes to the deformation of the capsules but also to their rotation. Therefore, increasing the value of the membrane viscosity leads to an increase in the dissipative effects on the surface due to viscous friction, which in turn reduces the energy available for deformation. If one deforms the capsule without using a flow but via external forces acting directly on the membrane (like the typical stretching experiment performed on RBCs by using optical tweezers (\cite{art:suresh2005connections})), the dependence of the steady-state value of the deformation on the membrane viscosity clearly disappears (\cite{guglietta2020effects,guglietta2021loading}). Additionally, in an elongational flow, where the rotation of the membrane is suppressed, the steady-state value of the deformation does not depend on the value of $\Bq$ (\cite{guglietta2021loading}). 

\corr{The steady-state values of the inclination angle $\theta$ are reported in Fig.~\ref{fig:d_and_t_vs_ca-single}(b). As expected, in the absence of membrane viscosity ($\Bq=0$), the inclination angle $\theta$ diminishes as a function of the capillary number, which is in good agreement with the results of \cite{aouaneStructureRheologySuspensions2021}). Upon introducing membrane viscosity, the inclination angle decreases, and intriguingly, exhibits a non-monotonic behaviour when $\Bq=50$.}

Fig.~\ref{fig:d_and_t_vs_ca-single}(c) shows that the loading time $\tl$ depends on both $\Ca$ and $\Bq$. In particular, on the one hand, it decreases when $\Ca$ increases, and seems to converge to a constant value. On the other hand, the increase of $\tl$ when the membrane viscosity increases is expected because of the viscous dissipation at the interface. The loading time $\tl$ depends on $\Bq$ even when we apply an elongational flow or perform a stretching experiment (\cite{guglietta2021loading}). This behaviour is opposite to that of the steady-state deformation value, which does not show such a dependence when only the membrane deformation is present. 

Fig.~\ref{fig:d_and_t_vs_ca-single}(d) depicts the frequency of the deformation oscillations $\tlcos$. It does not show a strong dependence on the membrane viscosity but only on the capillary number $\Ca$. This means that this characteristic time simply scales with the characteristic time of the flow, $\gammadot^{-1}$. 
The results for $\tl$ and $\tlcos$ are in qualitative agreement with results for a single RBC in simple shear flow (\cite{guglietta2021loading}).

\corr{In the literature, oscillations of the deformation have already been observed (\cite{art:yazdanibagchi13,art:lizhang19}) and also analytically predicted (\cite{barthesbiesel1985}). It is worth noticing that also droplets under simple shear flow exhibit such oscillations (\cite{art:gounley16}), meaning that they are not strictly related to the kind of the interface energy nor the presence of wrinkles -- since in that case surface tension acts and prevent any wrinkle appearing at the interface. Indeed, as explained by \cite{art:gounley16}, these oscillations appear when the flow time scale and the relaxation time scale differ significantly. } 

\corr{We also looked at the time evolution of the xz-component of the particle stress $\Sigma^p_{xz}$ given by the sum of the elastic and viscous contribution ($\Sigma^e_{xz}$ and $\Sigma^\nu_{xz}$, respectively), with the idea of bridging the micro- and macro-rheology by relating the loading time $\tl$ to the characteristic time the stress takes to reach the steady-state value (see Fig.~\ref{fig:sigma_vs_t-single}). When there is no membrane viscosity ($\Bq=0$), the particle stress is completely given by the elastic contribution; when $\Bq>0$, it is mainly dominated by the viscous contribution. Since $\Sigma^\nu_{xz}$ depends on the velocity gradient on the surface (see Eq.~\eqref{eq:bq-scriven-law}), it suddenly increases as soon as the shear flow starts, reducing thus the characteristic time of $\Sigma^p_{xz}$ almost to zero. This behaviour goes in the opposite direction with respect to what we observe for the loading time $\tl$, which is related to the deformation.}

Since the deformation as defined in Eq.~\eqref{eq:def} only contains information about the main axes in the shear plane, it does not provide a complete description of how the capsule is deforming in three-dimensional space. Therefore, we examined the three main radii $r_1, r_2$, and $r_3$ separately (Fig.~\ref{fig:r_vs_ca-single}, panels (a), (b), and (c), respectively.). The radii are normalised by the initial radius $R$, which is the capsule's radius at rest. In all three cases, the variation in the length of the radii $r_i$ ($i=1,2,3$) relative to their values at rest decrease as the membrane viscosity increases (as expected based on the measurements of the deformation). However, the most significant variation is seen in $r_1$ and $r_3$ (i.e., in the shear plane), while $r_2$ changes only slightly when $\Bq=0$ and is almost unchanged for $\Bq=50$.

\begin{figure}
\centering
\includegraphics[width=1.\linewidth]{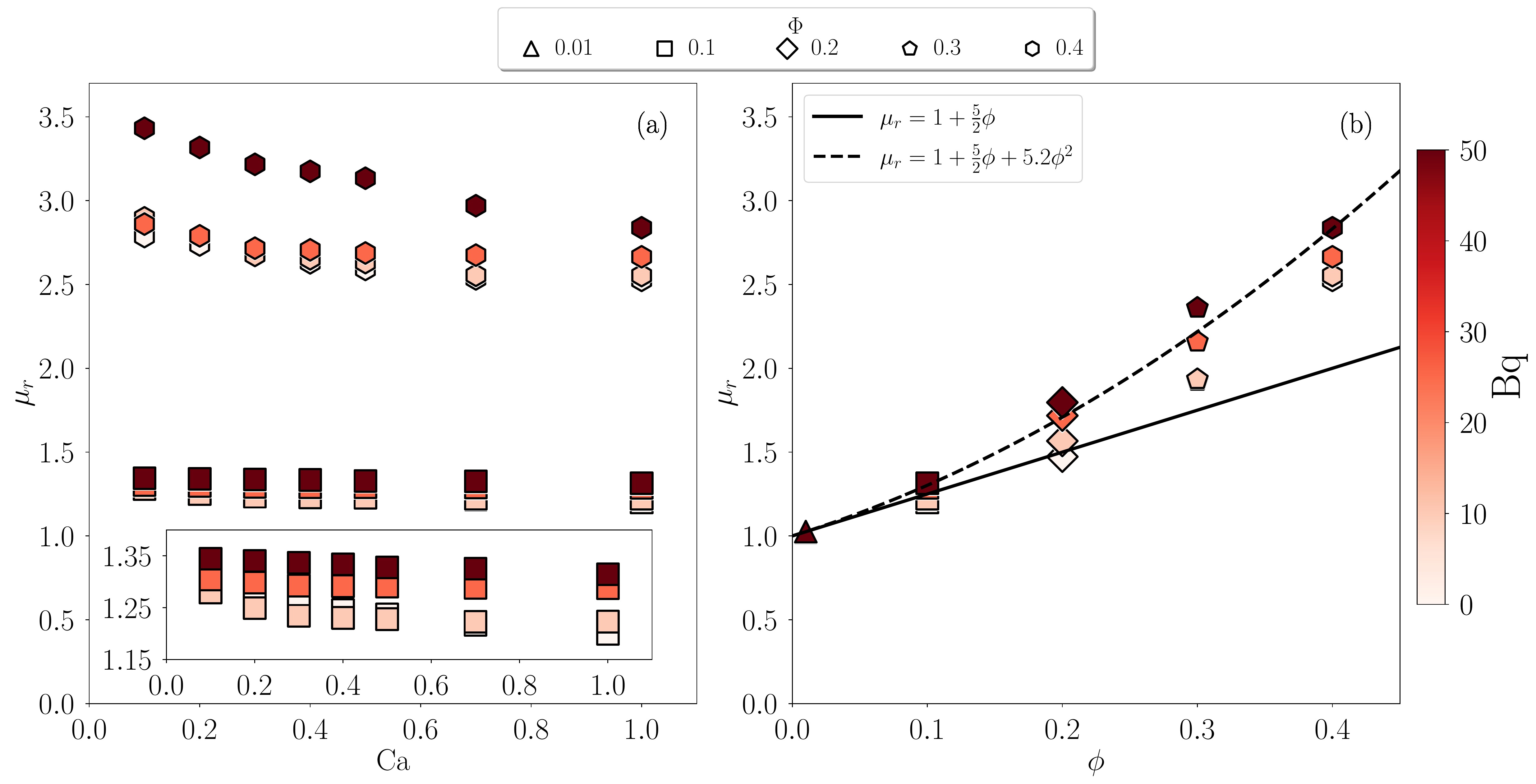}
\caption{\corr{Panel (a): $\mu_r$ as a function of $\Ca$ for different values of  $\Bq$ and $\phi=0.1$ and $0.4$. The inset shows magnified region around the $\phi=0.1$ data. Panel (b): $\mu_r$ as a function of $\phi$ for different values of $\Bq$ and $\Ca=1$. The solid and dashed lines are the theoretical predictions of  \cite{einstein1906neue} and \cite{batchelorDeterminationBulkStress1972}, respectively. }}\label{fig:mur}%
\end{figure}

\begin{figure}
\includegraphics[width=1.\linewidth]{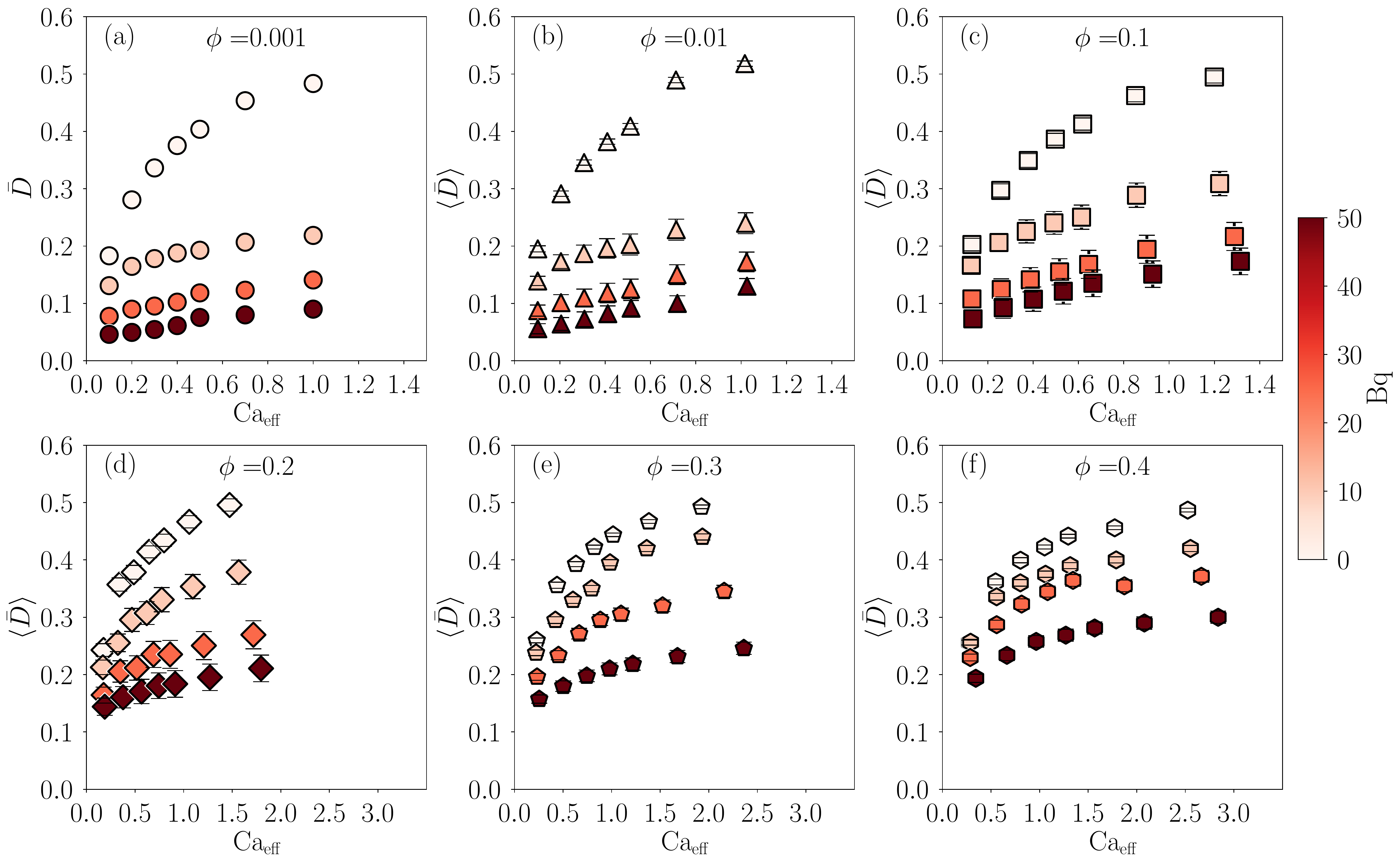}
\caption{\corr{Capsule-averaged steady-state deformation $\langle\bar{D}\rangle$  (see Eq.~\eqref{eq:fit}) as a function of $\Caeff$ for different values of $\phi$ (panel (a): $\phi = 0.001$; panel (b): $\phi = 0.01$;  panel (c): $\phi = 0.1$; panel (d): $\phi = 0.2$; panel (e): $\phi = 0.3$; panel (f): $\phi = 0.4$) and $\Bq$ 
($\Bq=0$, [\symbolsA]; 
$\Bq=10$, [\symbolsB]; 
$\Bq=25$, [\symbolsC]; 
$\Bq=50$, [\symbolsD]).  }
}\label{fig:d_vs_ca-susp}%
\end{figure}

\begin{figure}
\includegraphics[width=1.\linewidth]{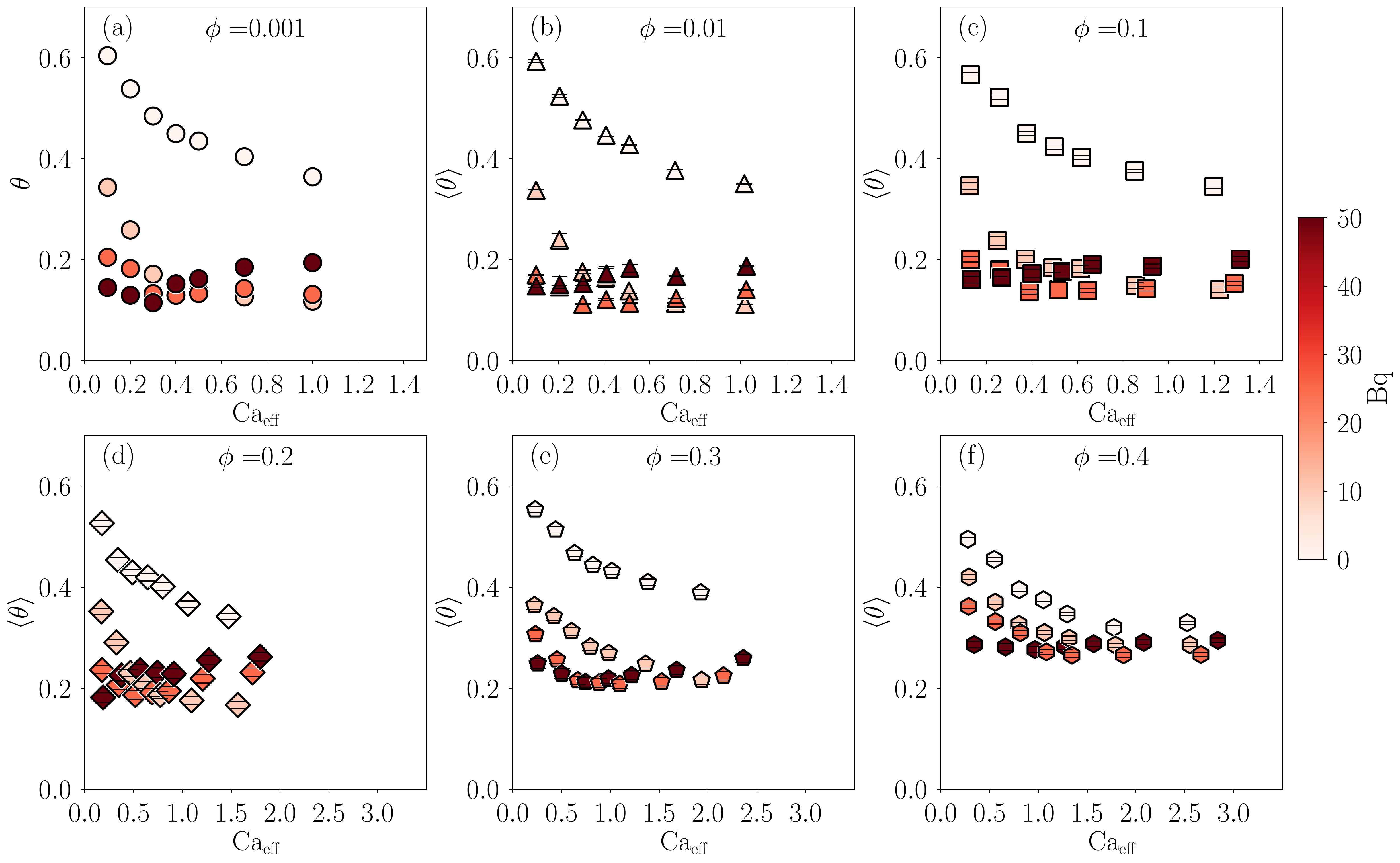}
\caption{\corr{Capsule-averaged steady-state inclination angle $\langle\bar{\theta}\rangle$  as a function of $\Caeff$ for different values of $\phi$ (panel (a): $\phi = 0.001$; panel (b): $\phi = 0.01$;  panel (c): $\phi = 0.1$; panel (d): $\phi = 0.2$; panel (e): $\phi = 0.3$; panel (f): $\phi = 0.4$) and $\Bq$ 
($\Bq=0$, [\symbolsA]; 
$\Bq=10$, [\symbolsB]; 
$\Bq=25$, [\symbolsC]; 
$\Bq=50$, [\symbolsD]).}
}\label{fig:theta_vs_ca-susp}%
\end{figure}

\begin{figure}
\includegraphics[width=1.\linewidth]{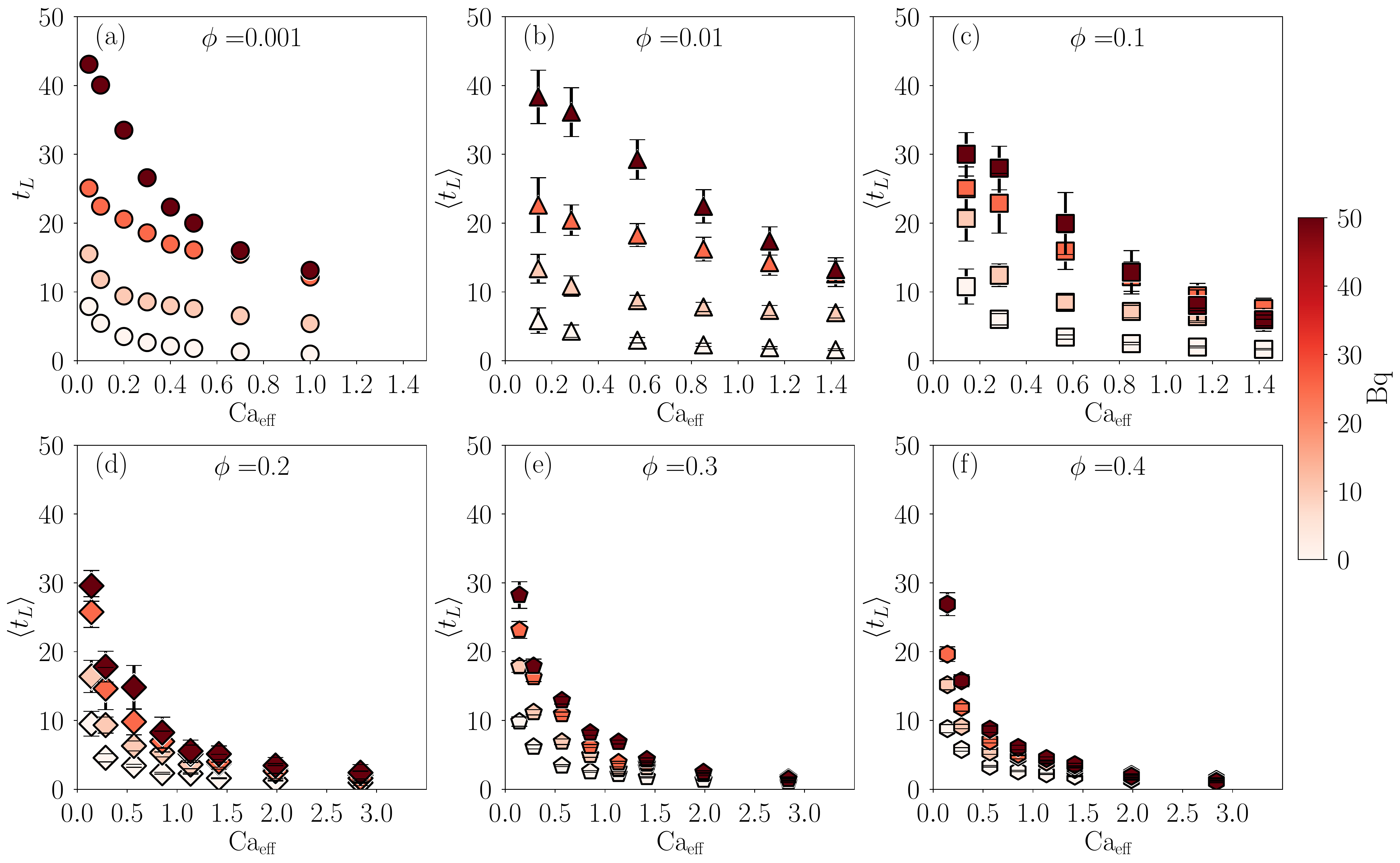}
\caption{\corr{Capsule-averaged loading time $\langle\tl\rangle$  (see Eq.~\eqref{eq:fit}) as a function of $\Caeff$ for different values of $\phi$ (panel (a): $\phi = 0.001$; panel (b): $\phi = 0.01$;  panel (c): $\phi = 0.1$; panel (d): $\phi = 0.2$; panel (e): $\phi = 0.3$; panel (f): $\phi = 0.4$) and $\Bq$ 
($\Bq=0$, [\symbolsA]; 
$\Bq=10$, [\symbolsB]; 
$\Bq=25$, [\symbolsC]; 
$\Bq=50$, [\symbolsD]).}
}\label{fig:tL_vs_ca-susp}%
\end{figure}

\begin{figure}
\centering
\includegraphics[width=.4\linewidth]{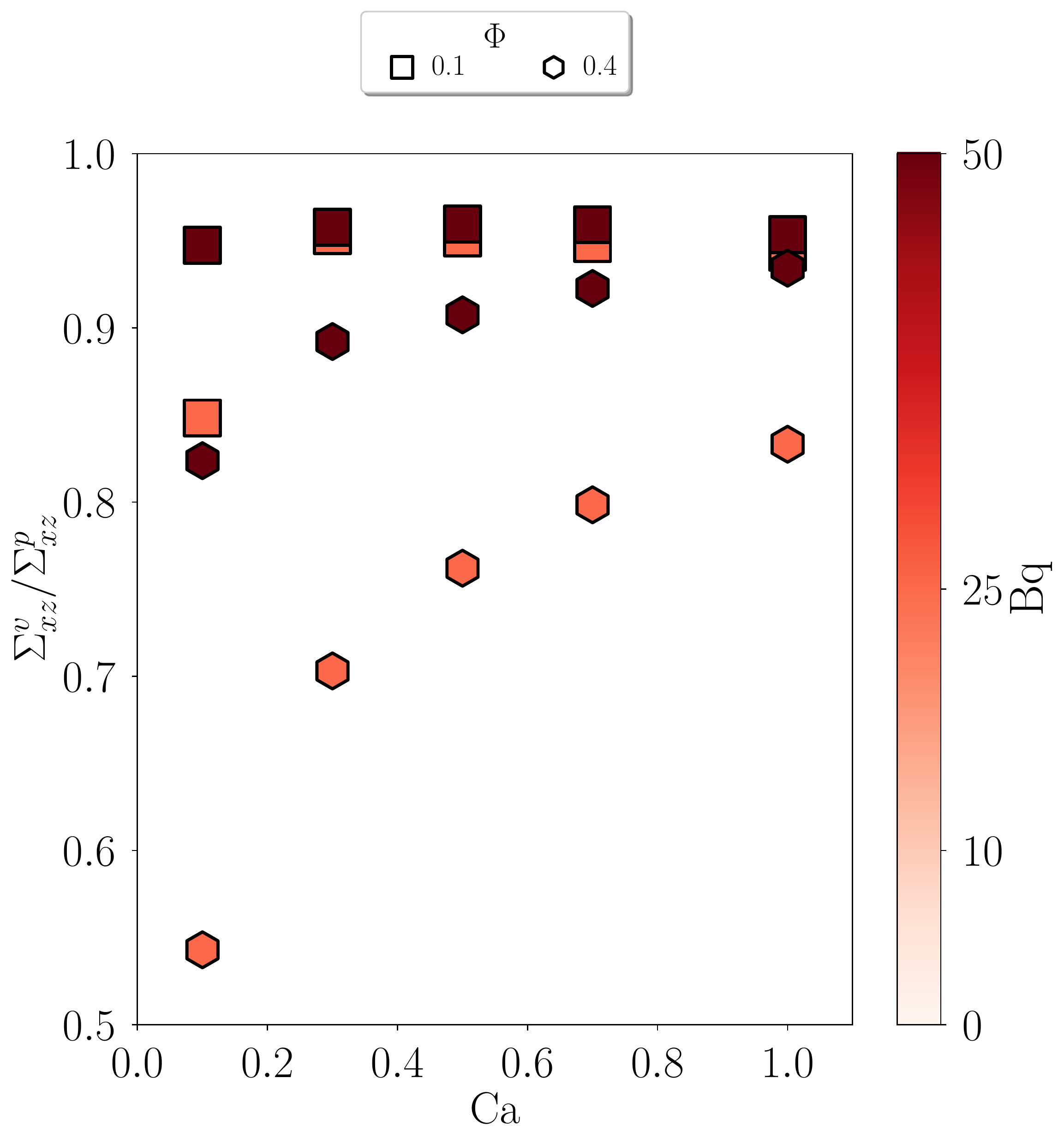}
\caption{\corr{Steady-state values of the ratio of the viscous contribution of the particle stress ($\Sigma^v_{xz}$) to the total stress of the particle ($\Sigma^p_{xz}=(\Sigma^v+\Sigma^e)_{xz}$) for $\Bq=25$, [\protect\squarec,\protect\exac] and $\Bq=50$, [\protect\squared,\protect\exad].} \label{fig:sigma-rel}}%
\end{figure}

\begin{figure}
\includegraphics[width=1.\linewidth]{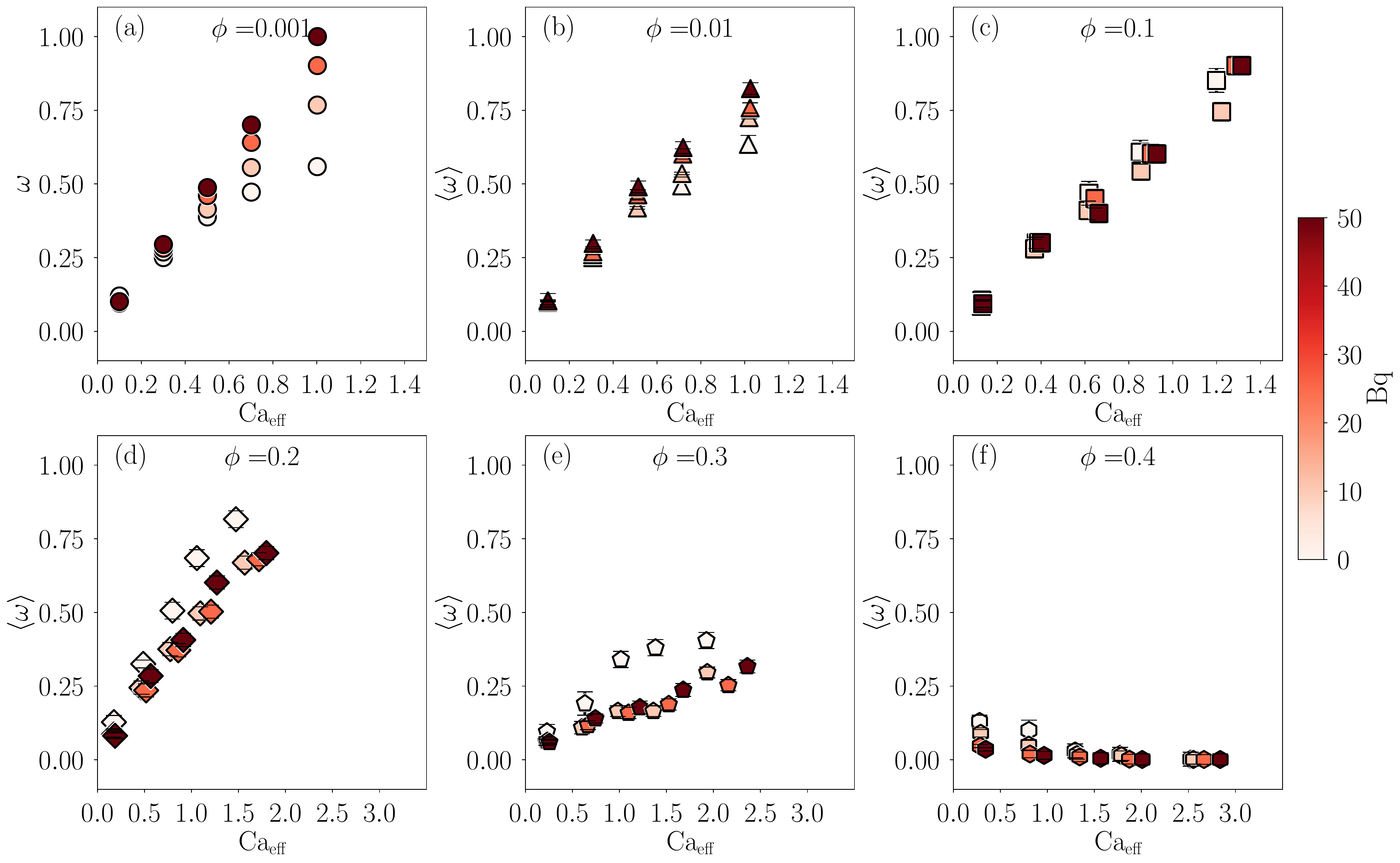}
\caption{\corr{Capsule-averaged frequency $\langle\tlcos\rangle$  (see Eq.~\eqref{eq:fit}) as a function of $\Caeff$ for different values of $\phi$ (panel (a): $\phi = 0.001$; panel (b): $\phi = 0.01$;  panel (c): $\phi = 0.1$; panel (d): $\phi = 0.2$; panel (e): $\phi = 0.3$; panel (f): $\phi = 0.4$) and $\Bq$
($\Bq=0$, [\symbolsA]; 
$\Bq=10$, [\symbolsB]; 
$\Bq=25$, [\symbolsC]; 
$\Bq=50$, [\symbolsD]).}
}\label{fig:tc_vs_ca-susp}%
\end{figure}

\begin{figure}
\includegraphics[width=1.\linewidth]{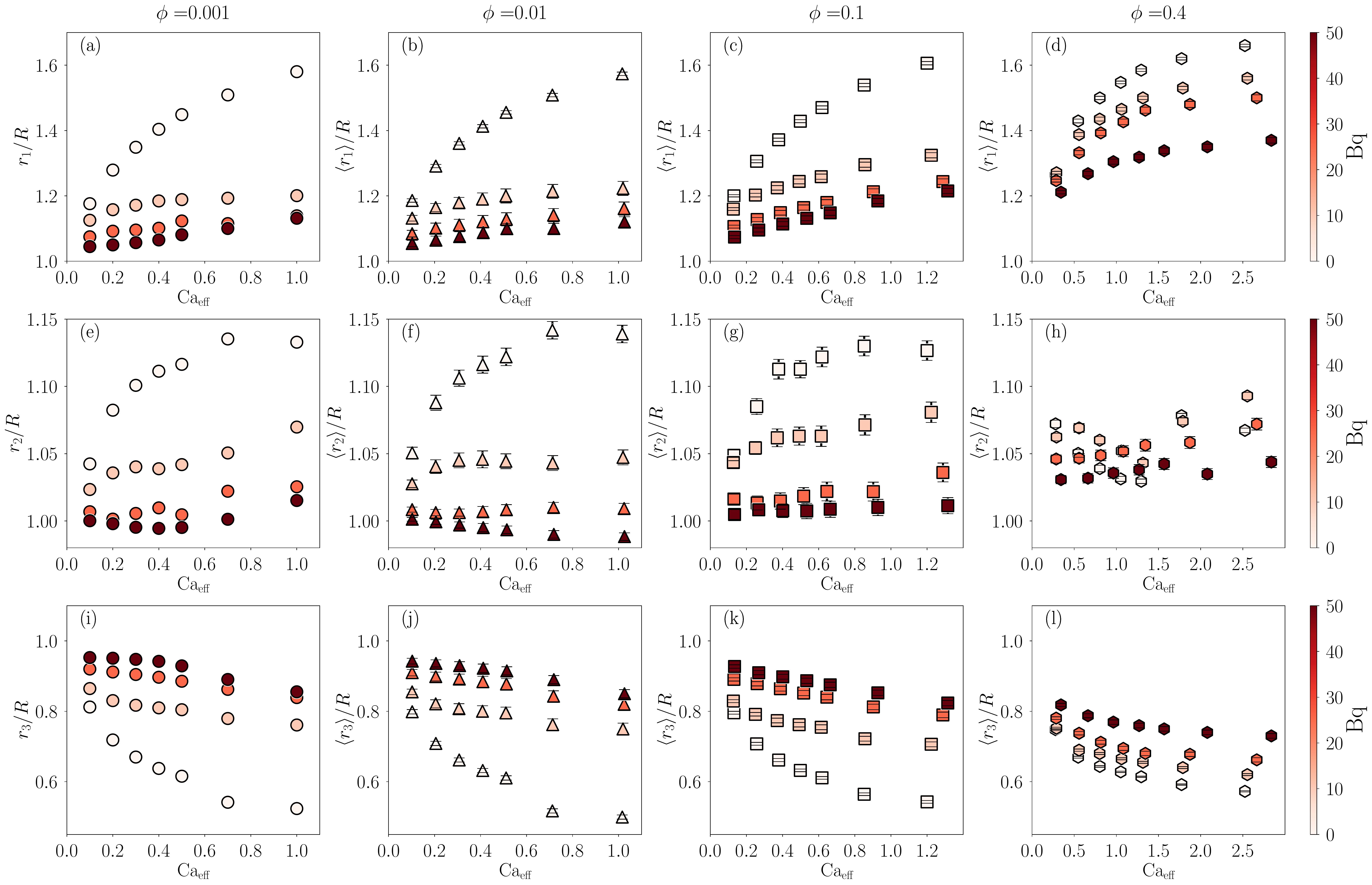}
\caption{\corr{The capsule-averaged lengths of the three main radii of the capsule $\langle r_1\rangle$ (panels (a)-(d)), $\langle r_2\rangle$ (panels (e)-(h)) and $\langle r_3\rangle$ (panels (i)-(l)), normalised with the radius of the spherical capsule at rest, $R$, as functions of $\Caeff$ for different values of $\phi$ (panels (a),(e),(i): $\phi = 0.001$; panels (b),(f),(j): $\phi = 0.01$; panels (c),(g),(k): $\phi = 0.1$; panels (d),(h),(l): $\phi = 0.4$) and the $\Bq$
($\Bq=0$, [\protect\circlea, \protect\trianglea, \protect\squarea, \protect\exaa]; 
$\Bq=10$, [\protect\circleb, \protect\triangleb, \protect\squareb, \protect\exab]; 
$\Bq=25$, [\protect\circlec, \protect\trianglec, \protect\squarec, \protect\exac]; 
$\Bq=50$, [\protect\circled, \protect\triangled, \protect\squared, \protect\exad]). }
}\label{fig:r_vs_ca-susp}%
\end{figure}

\subsection{Suspensions}\label{sec:res_def_susp}
We consider the same numerical setup as before, but now we increase the number of capsules $N$ up to 400, corresponding to an increase of the volume fraction $\phi$ up to 0.4. We introduce the capsule-averaged quantities represented by 
\begin{equation}
    \langle A \rangle = \frac{1}{N}\sum_i A_i\ ,
\end{equation}
where the sum runs over the number of particles $N$ and $A_i$ is a general observable measured for the $i-$th capsule (such as the steady-state value of the deformation $\bar{D}$, the loading time $\tl$, the radius $r_i$, etc.). The data reported in this section are provided with error bars, which are calculated from the standard deviation normalised with $\sqrt{N}$.

Fig.~\ref{fig:sketch-susp} shows some steady-state configurations for three different values of $\phi$ (columns) and two values of $\Bq$ (rows). Data refer to $\Ca=0.1$.
It is interesting to observe that wrinkles do not appear on the surface when $\Bq=0$, whereas they are visible for $\Bq=50$. However, in the latter case, the volume fraction seems to play a role: indeed, while the cases with $\phi=0.01$ and $\phi=0.1$ show just a few particles with small wrinkles (panels (d) and (e), respectively), the most dense case (panel (f)) shows more pronounced wrinkles on more particles.  

\corr{We want to study the transient dynamics of the system and compare results for different values of the volume fraction $\phi$. To make the comparison as fair as possible, we initialise the system without membrane pre-stress also for the dense case, since that is the case for the dilute suspensions. To reach high volume fractions ($\phi>0.1$) without deforming the capsules, we initialised the system in an fcc crystal configuration (see Fig.~\ref{fig:sketch-susp}, panels (c) and (f)).}

\corr{Before analysing the same quantities studied in the single particle case, we investigate the rheological properties of the suspension. We consider the xz-component of the stress of the capsule $\Sigma_{xz}^p$ and its elastic and viscous components ($\Sigma_{xz}^e$ and $\Sigma_{xz}^\nu$, respectively). In Fig.~\ref{fig:mur}, we report the relative viscosity $\mur =\frac{\mu_s}{\mu} =1 + \frac{\Sigma_{xz}^p}{\gammadot\mu}$, where $\mu_s$ is the effective viscosity of the suspension. For very dilute suspensions, ($\phi = 0.01$) the relative viscosity $\mur\approx 1$. Upon increasing the volume fraction $\phi$, we observe an expected increase in $\mur$. In  Fig.~\ref{fig:mur}, panel (a), we report the $\mur$ as a function of the capillary number $\Ca$, for different values of $\Bq$. For the sake of clarity, we report only data for $\phi=0.1$ and $\phi=0.4$. In both cases, we observe an increase of the relative viscosity with $\Bq$. To better appreciate this dependency in the $\phi=0.1$ case, this is magnified in the inset, showing that the shear-thinning behaviour is present regardless of the value of $\Bq$, but is more pronounced for $\Bq=50$. In  Fig.~\ref{fig:mur}, panel (b), we report $\mur$ as a function of the volume fraction $\phi$ for different values of $\Bq$. Again, to improve the readability of the plot, we selected data for the highest value of capillary number only, $\Ca=1$. We also report the theoretical predictions of $\mur(\phi)$ according to \cite{einstein1906neue} ($\mur=1+\frac{5}{2}\phi$, solid black line) and   \cite{batchelorDeterminationBulkStress1972} ($\mur=1+\frac{5}{2}\phi+5.2\phi^2$, dashed black line), which hold for suspensions of hard spheres in the dilute and semi-dilute approximations, respectively. It is worth noticing that the cases with small values of $\Bq$ differ from the prediction computed by \cite{batchelorDeterminationBulkStress1972}, but still show a quadratic behaviour, while the data for high values of $\Bq$ are closer to values predicted by the theory for hard spheres. The reason  may be that a high membrane viscosity reduces the deformation of the capsule, making them a better approximation of hard spheres, at least from a geometrical point of view. The change in relative viscosity calls for a redefinition of the capillary number.  When $\phi$ increases, the viscosity of the suspension increases too (as shown in Fig.~\ref{fig:mur}). Therefore, we introduce the effective capillary number $\Caeff$, which accounts for the viscosity of the suspension $\mu_s$:
\begin{equation}\label{eq:caeff}
    \Caeff = \frac{\gammadot R \mu_s}{k_s}= \Ca\ \mur\ .
\end{equation}}

In Fig.~\ref{fig:d_vs_ca-susp}, the capsule-averaged steady-state deformation is reported as a function of the \corr{$\Caeff$} for different values of $\Bq$ and $\phi$. The data for the single capsule ($\phi=0.001$) are also reported for comparison (Fig.~\ref{fig:d_vs_ca-susp}, panel (a)). 
As already observed for elastic capsules in the absence of membrane viscosity, our data shows that the capsule-averaged steady-state deformation $\langle\bar{D}\rangle$ slightly increases with increasing $\phi$ (\cite{aouaneStructureRheologySuspensions2021}). 
It is interesting to compare panels (a) and (f), which are the two extreme cases we simulated (i.e., $\phi=0.001$ and $0.4$, respectively). We observe that in absence of membrane viscosity ($\Bq=0$), $\langle\bar{D}\rangle(\phi=0.4)$ is about $5-10$\% higher than $\langle\bar{D}\rangle(\phi=0.001)$, while when $\Bq=50$, there is an increase of about 250\%. This suggests a weaker effect of the membrane viscosity in reducing the deformation for higher values of $\phi$. This general trend can be observed in Fig.~\ref{fig:d_vs_ca-susp} for all the reported values of $\phi$. We note that $\langle\bar{D}\rangle$ increases when $\phi$ increases (from panel (a) to (f)) at $\Bq=50$, but this difference in $\langle\bar{D}\rangle$ shrinks when $\Bq$ is smaller. 

\corr{The values of the inclination angle $\theta$ as a function of $\Caeff$ are reported in Fig.~\ref{fig:theta_vs_ca-susp}. For volume fraction up to $\phi=0.3$, the results are very similar to the single-capsule case, with a slight increase for $\Bq\ne 0$. However, in the most dense case simulated, the inclination angle is slightly reduced (with respect to the single capsule case) in absence of membrane viscosity, while it is increased in the other cases. It is interesting to note the  collapse of $\theta$ for high values the effective capillary number, $\Caeff>1.5$.}

Regarding the capsule-averaged loading time $\langle\tl\rangle$, depicted in Fig.~\ref{fig:tL_vs_ca-susp}, we observe again that the volume fraction $\phi$ mitigates the effect of the presence of the membrane viscosity, especially at increasing values of the capillary number. In fact, the apparent increase of  $\langle\tl\rangle$ with $\Bq$ for $\phi\le 0.01$ (panels (a)-(c)) is not present for higher values of $\phi$ (panels (d)-(f)). Furthermore, it is worth noting that $\langle\tl\rangle$ shows a slight dependence on the volume fraction $\phi$ for $\Bq=0$, and the $\phi=0.001$ and $\phi=0.4$ data superpose almost perfectly. The dependence of $\langle\tl\rangle$ on $\phi$ and $\Bq$ is even more evident for small values of the capillary number $\Ca$ (close to the linear response), i.e., when focusing on the intrinsic properties of the membrane: for the volume fraction $\phi\ge 0.1$, $\langle\tl\rangle$ still shows a dependence on $\Bq$, but if the capillary number $\Ca$ increases, the data tend to collapse on the same curve. This means that, for suspensions with a concentration $\phi\ge 0.1$ and for high values of \corr{the effective capillary number $\Caeff$}, the effect of membrane viscosity almost disappears. 
The origin of the reduction of the effect of membrane viscosity with volume fraction increase can be traced to the viscous tensor defined in Eq.~\eqref{eq:bq-scriven-law}: while the elastic contribution depends only on the geometry (i.e., the deformation) of the capsule, the viscous tensor depends only on the surface velocity gradient $\boldsymbol{\nabla}^{\boldsymbol{S}}\vec{u}^{\boldsymbol{S}}$. Therefore, when the volume fraction $\phi$ increases, the strain tensor $\vec{e}$ (see Eq.~\eqref{eq:strain}) decreases, and the effect of the membrane viscosity becomes smaller. 
\corr{In Fig.~\ref{fig:sigma-rel}, we report the ratio $\Sigma^v_{xz}/\Sigma^p_{xz}$ as a function of the capillary number $\Ca$ for two values of volume fraction ($\phi=0.1$ and $\phi=0.4$) and for two values of $\Bq$ ($\Bq=25$ and $\Bq=50$). We observe a reduction of the contribution given by $\Sigma^v_{xz}$ when the volume fraction increases, while for dilute suspensions the ratio  $\Sigma^v_{xz}/\Sigma^p_{xz}$ is close to 1. This suggests that, when the volume fraction increases, the viscous dissipation reduces and the energy left contributes to the elastic deformation.}

Concerning the capsule-averaged frequency of the oscillations $\omega$, we observe that there is a weak dependence on $\Bq$ for volume fractions up to $\phi=0.1$ (see Fig.~\ref{fig:tc_vs_ca-susp}, panels (a)-(c)); however, for $\phi>0.1$ (panels (d)-(f)), the oscillations of the deformation disappear, and therefore $\omega$ goes to zero at large \corr{$\Caeff$}. This may be due to the \corr{collisions (i.e., strong capsule-capsule interactions) that do} not allow the deformation of the capsules to oscillate freely. 
\corr{We also looked at the deformation of some capsules for the most dense case simulated ($\phi=0.4$) and we observed that the deformation shows small and noisy fluctuations around the average: again, these oscillations can be attributed to the capsule-capsule collisions. By looking at the deformation for some capsules in the suspensions, we also checked that the behaviour of the capsule-averaged deformation well reflects the one of the single capsules, making Eq.~\eqref{eq:fit} still good to estimate $\tl$ and $\omega$. To further confirm the goodness of the fitting procedure, the reader can look at the error bars reported in Figs.~\ref{fig:d_vs_ca-susp}-\ref{fig:tL_vs_ca-susp}.} 

As presented in the previous section for the single capsule, in Fig.~\ref{fig:r_vs_ca-susp} we show the capsule-averaged values of the normalised radii $\langle r_1\rangle/R$, $\langle r_2\rangle/R$ and $\langle r_3\rangle/R$ (panels (a)-(d), (e)-(h) and (i)-(l), respectively). We observe that, at a given value of the volume fraction, the membrane viscosity clearly reduces the deformation of the three radii. The effect of the volume fraction becomes important for $\phi >0.1$, that is, when capsules start to interact with each other. Even when the volume fraction increases, most of the deformation occurs in the shear plane (i.e., $r_2$ is less affected than $r_1$ and $r_3$). The effect of the volume fraction becomes prominent for $\phi>0.1$: indeed, for all the values of $\Bq$ we have simulated, when $\phi=0.4$ the radii $r_1$ and $r_3$ (panels (d) and (l)) are different if compared with the cases $\phi\le 0.1$. This might be due to the strong capsule-capsule interaction when $\phi=0.4$, confirming again that the effect of membrane viscosity reduces for high values of the volume fraction. 
Concerning the deformation in the vorticity direction, $r_2$, it is  $\sim 10\%$ for $\Bq=0$ and $\lesssim 5\%$ for $\Bq>0$. While $r_2$ shows a clear hierarchy in $\Bq$ for $\phi<0.4$, a more complex behaviour appears when $\phi=0.4$. However, we are facing very small deformations (less than $5\%$), which means that the length of $r_2$ changes by about $0.4\Delta x$. We conclude that the deformation in the vorticity direction is in general small, especially when we increase the volume fraction. To provide a more quantitative and precise investigation for the behaviour of $r_2$, one should perform simulations with larger capsules (and therefore with a more resolved mesh); however, such a detailed study on the deformation in the vorticity direction goes beyond the scope of this work.

\section{Conclusions}\label{sec:summary}
In this study, we performed a parametric investigation of the impact of membrane viscosity on the transient dynamics of suspensions of viscoelastic spherical capsules for different values of the volume fraction $\phi$. 
To achieve this, we performed numerical simulations using the IB-LB method. 
Our results indicate that the effect of membrane viscosity, as measured by the dimensionless Boussinesq number $\Bq$, strongly impacts the dynamics of a single capsule. However, this effect is diminished as the volume fraction $\phi$ increases. The comparison between the single-capsule case ($\phi=0.001$) and the most-dense case simulated ($\phi=0.4$) revealed that while the capsule-averaged deformation $\langle\bar{D}\rangle$ is greatly affected by the presence of membrane viscosity, the capsule-averaged loading time $\tl$ does not show a strong dependence on $\Bq$ when $\phi=0.4$.
We can therefore conclude that, for the flow conditions simulated in this work (i.e., $\Rey\sim 0.01$ and $\Ca\in[0.05,1]$, as outlined in Tab.~\ref{tab:values}), the  membrane viscosity does not significantly affect the characteristic time when the volume fraction is high enough, but it still has a substantial impact on the deformation.

In the future it will be valuable to investigate the dynamics of both dilute and dense suspensions flowing through small channels. The interaction between membrane viscosity and confinement is yet to be studied in this context. Additionally, it would be of interest to study the effect of membrane viscosity on different geometries and membrane models, with a focus on red blood cells as an example.



\section{Acknowledgements}{This work has received financial support from the Deutsche Forschungsgemeinschaft (DFG, German Research Foundation) – Project-ID 431791331 – SFB 1452 ``Catalysis at liquid interfaces'' and research unit FOR2688 ``Instabilities, Bifurcations and Migration in Pulsatile Flows'' (Project-ID 417989464).  
This work was supported by the Italian Ministry of University and Research (MUR) under the FARE programme, project ``Smart-HEART''.
The authors gratefully acknowledge the Gauss Centre for Supercomputing e.V. (\url{www.gauss-centre.eu}) for funding this project by providing computing time through the John von Neumann Institute for Computing (NIC) on the GCS Supercomputer JUWELS (\cite{JUWELS}) at Jülich Supercomputing Centre (JSC).
}

\printbibliography
\end{document}